%% file: main.tex
\documentclass[runningheads,envcountsame]{llncs}

\makeatletter \RequirePackage[bookmarks,unicode,colorlinks=true]{hyperref}%
\def\@citecolor{blue}%
\def\@urlcolor{blue}%
\def\@linkcolor{blue}%

\def\orcidID#1{\smash{\href{http://orcid.org/#1}{\protect\raisebox{-1.25pt}{\protect\includegraphics{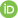}}}}}
\makeatother

\usepackage{xcolor}
\usepackage{hyperref}
\usepackage{mathtools}
\usepackage{multirow}
\usepackage{stmaryrd}
\usepackage{cancel}
\usepackage{amsmath,amssymb}
\usepackage{thmtools, thm-restate}
\usepackage{nccmath}
\usepackage[shortlabels]{enumitem}
\usepackage{microtype}
\usepackage{pbox}
\usepackage[ruled, vlined, linesnumbered]{algorithm2e}
\hypersetup{
	pdftitle={Targeting Completeness: Using Closed Forms for Size Bounds of Integer Programs}, colorlinks=true, linkcolor=blue, citecolor=olive, filecolor=magenta, urlcolor=cyan
}
\usepackage{todonotes}
\usepackage{placeins}
\usepackage{tikz}
\usepackage{caption}
\usepackage{subcaption}
\usepackage{mymatrix}

\usetikzlibrary{shapes,calc,arrows,automata,decorations.pathmorphing}
\usepackage[capitalize,nameinlink]{cleveref}
\usepackage{IEEEtrantools}

\newenvironment{myproof}{
	\noindent{\it Proof.}
}{\qed
	\medskip
}

\usepackage[location=appendix,prependtoappendix=true]{moveproofs}
\input{commands}

\SetKwInOut{Input}{input}
\SetKwInOut{Output}{output}

   \usepackage{cite}

\title{Targeting Completeness: Using Closed Forms for Size Bounds of Integer
  Programs\thanks{funded by the Deutsche Forschungsgemeinschaft (DFG, German Research
    Foundation) - 235950644 (Project GI 274/6-2)}}
\titlerunning{Targeting Completeness: Using Closed Forms for Size Bounds}
\author{Nils Lommen\orcidID{0000-0003-3187-9217} \and Jürgen
  Giesl\orcidID{0000-0003-0283-8520}}
\authorrunning{Nils Lommen and Jürgen Giesl}
\institute{LuFG Informatik 2, RWTH Aachen University, Aachen, Germany}

\begin{document}
\allowdisplaybreaks \maketitle
\input{abstract}

\input{introduction}

\input{sizebound_loops}
\input{solvable_loops}

\input{integer_programs}

\input{complexity}

\input{conclusion}

  \bibliographystyle{splncs04}
  \newcommand{\noopsort}[1]{}
  \input{mainArxiv.bbl}

\end{document}

%% file: commands.tex
\makeatletter \newcommand*\bigcdot{\mathpalette\bigcdot@{.4}}
\newcommand*\bigcdot@[2]{\mathbin{\vcenter{\hbox{\scalebox{#2}{$\m@th#1\bullet$}}}}}
\makeatother

\renewcommand{\emptyset}{\varnothing}

\newcommand{\tool}[1]{\textsf{#1}}
\newcommand{\emphit}[1]{\textit{#1}}

\newcommand{\myvec}[1]{\ensuremath{
		\begin{sbmatrix}
			#1
		\end{sbmatrix}
	}}
\newcommand{\pvec}[1]{\vec{#1}\mkern2mu\vphantom{#1}}
\newcommand{\dontprint}[1]{}
\newcommand{\xvec}{\vec{x}}

\newcommand{\eat}[1]{}

\newcommand{\NN}{\mathbb{N}}
\newcommand{\ZZ}{\mathbb{Z}}

\newcommand{\NNC}{\overline{\mathbb{N}}}
\newcommand{\QQ}{\mathbb{Q}}

\renewcommand{\AA}{\mathbb{A}}

\newcommand{\PP}{\mathcal{P}}

\newcommand{\var}{\texttt}

\newcommand{\true}{\var{true}}

\newcommand{\cl}[1]{\normalfont{\texttt{cl}^{#1}}}
\newcommand{\clExp}[2]{\normalfont{\texttt{cl}^{#1}_{#2}}}

\newcommand{\sgn}{\mathrm{sgn}}
\newcommand{\location}{\ell}
\newcommand{\state}{\sigma}
\newcommand{\initial}{\sigma_0}
\newcommand{\State}{\Sigma}
\newcommand{\update}{\eta}

\newcommand{\guard}{\varphi}

\newcommand{\landau}{\mathcal{O}}

\newcommand{\entry}{\mathcal{E}}

\newcommand{\Size}{{\mathcal{SB}}}
\newcommand{\size}{\texttt{sb}}

\newcommand{\loc}{{\mathcal{RB}_{\TSet'_>}}}
\newcommand{\glo}{{\mathcal{RB}}}

\DeclareMathOperator{\rc}{rc}

\newcommand{\pret}{r}
\newcommand{\actt}{t}

\newcommand{\BoundSet}{\mathcal{B}}

\newcommand{\FormulaSet}{\mathcal{F}}

\newcommand{\TSet}{\mathcal{T}}
\newcommand{\VSet}{\mathcal{V}}

\newcommand{\SSet}{\mathcal{S}}
\newcommand{\LSet}{\mathcal{L}}

\newcommand{\IntLoop}{(\guard, \update)}
\newcommand{\IntProgram}{(\VSet,\LSet,\location_0,\TSet)}

\newcommand{\braced}[1]{\lbrace #1 \rbrace}

\newcommand{\timeboundterm}{
	\sup \braced{ n \in \NN \mid \exists \, (\location', \state').\; (\location_0, \state_0) \; (\rightarrow^*_{\TSet} \circ \rightarrow_t)^n \; (\location', \state') }
}

\newcommand{\sizeboundterm}{
	\braced{ |\state'(v)| \mid \exists\, \location' \in \LSet.
		\; (\location_0, \state_0) \; (\rightarrow^* \circ \rightarrow_\actt) \; (\location', \state')}
}
\newcommand{\sizeboundtermx}{
	\braced{ |\state'(x)| \mid \exists\, \location' \in \LSet.
		\; (\location_0, \state_0) \; (\rightarrow^* \circ \rightarrow_\actt) \; (\location', \state')}
}

\newcommand{\chain}{\star}

\newcommand{\compconj}[1]{%
	\overline{#1}%
}

\crefname{definition}{Def.}{Def.}
\crefname{example}{Ex.}{Ex.}
\crefname{counterexample}{Counterex.}{Counterex.}
\crefname{appendix}{App.}{App.}
\crefname{ex}{Ex.}{Ex.}
\crefname{theorem}{Thm.}{Thm.}
\crefname{lemma}{Lemma}{Lemmas}
\crefname{remark}{Rem.}{Rem.}
\crefname{section}{Sect.}{Sect.}
\crefname{subsection}{Sect.}{Sect.}
\crefname{subsubsection}{Sect.}{Sect.}
\crefname{line}{Line}{Lines}
\crefname{corollary}{Cor.}{Cor.}
\crefname{figure}{Fig.}{Fig.}
\crefname{enumi}{}{}
\crefname{algorithm}{Alg.}{Alg.}

\newcommand{\KoAT}[0]{\tool{KoAT}}
\DeclarePairedDelimiter\ceil{\lceil}{\rceil}

\newcommand{\im}{\mathrm{i}}

%% file: abstract.tex
\begin{abstract}
	We  present a new procedure to infer \emph{size bounds}
	for integer programs automatically.
	Size bounds are important for the deduction of bounds on the runtime complexity or in general, for the resource analysis of programs.
        We show that our technique is \emph{complete} (i.e., it always computes finite size bounds) for a subclass of loops, possibly with non-linear arithmetic.
	Moreover, we present a novel approach to combine and inte\-grate this complete
        technique into an incomplete approach
        to infer size and\linebreak
	runtime bounds of general integer programs.
	We prove completeness of our integration for an important subclass of integer programs.
        We implemen\-ted our new algorithm in the automated complexity analysis tool \KoAT{}
	to\linebreak
	evaluate its power, in particular on programs with non-linear arithmetic.
\end{abstract}

%% file: introduction.tex
\section{Introduction}

There are numerous incomplete approaches for automatic resource analysis of programs, e.g., \cite{ben-amram2017MultiphaseLinearRankingFunctions,albert2019ResourceAnalysisDriven,sinn2017ComplexityResourceBound,brockschmidt2016AnalyzingRuntimeSize,cofloco2,Festschrift,ramlpopl17,lopez18,carbonneaux2015CompositionalCertifiedResource,albert2012CostAnalysisObjectoriented}.
However, also many complete techniques to decide termination, analyze runtime complexity, or study memory consumption for certain classes of programs have been developed, e.g., \cite{dblp:conf/cav/tiwari04,dblp:conf/cav/braverman06,frohn2019TerminationTriangularInteger,ben-amram2019TightWorstCaseBounds,hosseini2019TerminationLinearLoops,frohn2020TerminationPolynomialLoops,hark2020PolynomialLoopsTermination,lommen2022AutomaticComplexityAnalysis,ben-amram2016FlowchartProgramsRegular,ben-amram2008LinearPolynomialExponential,xuSymbolicTerminationAnalysis2013}.
In this paper, we present a procedure to compute \emph{size bounds} which indicate how large the absolute value of an integer variable may become.
In contrast to other complete procedures for the inference of size bounds which are based on fixpoint computations \cite{ben-amram2019TightWorstCaseBounds,ben-amram2008LinearPolynomialExponential}, our technique can also handle (possibly negative) constants and exponential size bounds.
Similar to our earlier paper \cite{lommen2022AutomaticComplexityAnalysis}, we embed a procedure which is \emph{complete} for a subclass of loops (i.e., it computes finite size bounds for all loops from this subclass) into an incomplete approach for general integer programs \cite{Festschrift,brockschmidt2016AnalyzingRuntimeSize}.
In this way, the power of the incomplete approach is increased significantly, in particular for programs with non-linear arithmetic.
However, in the current paper we tackle a completely different problem than in \cite{lommen2022AutomaticComplexityAnalysis}
(and thus, the actual new contributions are also completely different), because in \cite{lommen2022AutomaticComplexityAnalysis}
we embedded a complete technique in order to infer runtime bounds, whereas now we integrate a novel technique in order to infer size bounds.
As an example, we want to determine bounds on the absolute values of the variables during (and after) the execution of the following loop.
\begin{equation}
	\label{WhileExample}
	\hspace*{-.1cm}	\textbf{while }\!(x_3 > 0)\!\textbf{ do }\! (x_1, x_2, x_3,x_4) \!\leftarrow\! (3\cdot x_1 + 2\cdot x_2, -5\cdot x_1 -3\cdot x_2, x_3 - 1, x_4 + x_3^2)
\end{equation}

We introduce a technique to compute size bounds for loops which admit a closed form,
i.e., an expression which corresponds to applying the loop's update $n$ times.
Then we over-approximate the closed form to obtain a non-negative, weakly monotonically increasing function.
For instance, a closed form for $x_3$ in our example is $x_3 - n$, since the value of $x_3$ is decreased by $n$ after $n$ iterations.
The (absolute value of this) closed form can be over-approximated by $x_3 + n$, which is monotonically increasing in all variables.
Finally, each occurrence of $n$ is substituted by a runtime bound for the loop.
Clearly, \eqref{WhileExample} terminates after at most $x_3$ iterations.
So if we substitute $n$ by the runtime bound $x_3$ in the over-approximated closed form $x_3 + n$, then we infer the linear bound $2\cdot x_3$ on the size of $x_3$.
Due to the\linebreak
restriction to weakly monotonically increasing over-approximations, we can plug in\linebreak
any over-approximation of the runtime and do not necessarily need exact bounds.

\paragraph*{Structure}
We introduce our technique to compute size bounds by closed forms in \cref{sect:size_bounds_closed_forms} and show that it is complete for a subclass of loops 
in \cref{sect:solvable_loops}.
Afterwards in \cref{sect:integer_programs}, we incorporate our novel technique into the incomplete setting of general integer programs.
In \cref{sect:complexity} we demonstrate how size bounds are used in automatic complexity analysis and study completeness for classes of general programs.
In \cref{sect:conclusion}, we conclude with an experimental evaluation of our
implementation in the tool \KoAT{} and discuss related work.
All proofs can be found in \cref{app:proofs}.

%% file: sizebound_loops.tex
\section{Size Bounds by Closed Forms}
\label{sect:size_bounds_closed_forms}
In this section, we present our novel technique to compute size bounds for loops by closed forms in \Cref{thm:size_bounds_closed_form}.
We start by introducing the required preliminaries.
Let $\VSet=\braced{x_1, \ldots, x_d}$ be a set of variables.
$\FormulaSet(\VSet)$ is the set of all \emph{formulas} built from inequations $p > 0$ for
polynomials
$p\in\QQ[\VSet]$,
$\wedge$, and $\vee$.
A \emph{loop} $\IntLoop$ consists of a guard $\guard\in\FormulaSet(\VSet)$ and an update $\update: \VSet \rightarrow \ZZ[\VSet]$ mapping variables to polynomials.
A \emph{closed form} $\cl{x_i}$ (formally defined in \Cref{Closed Form Def} below)
is an expression in $n$ and in
the (initial values of the) variables $x_1,\ldots,x_d$ which corresponds to the value of
$x_i$ after iterating the loop $n$ times.
For our purpose we only need closed forms which hold for all $n \geq n_0$ for some fixed $n_0\in\NN$.
Moreover, we restrict ourselves to closed forms which are so-called normalized poly-exponential expressions \cite{frohn2019TerminationTriangularInteger}.
Nonetheless, our procedure works for any closed form expression with a finite number of arithmetic operations (i.e., the number of operations must be independent of $n$).
We extend the application of functions like $\update: \VSet \rightarrow \ZZ[\VSet]$ also to polynomials, vectors, and formulas, etc., by replacing each variable $v$ in the expression by $\update(v)$.
So in particular, $(\update_2 \circ \update_1)(x) = \update_2(\update_1(x))$ stands for the polynomial $\update_1(x)$ in which every variable $v$ is replaced by $\update_2(v)$.
Moreover, $\update^n$ denotes the $n$-fold application of $\update$.

We call a function $\state: \VSet \rightarrow \ZZ$ a \emph{state}.
By $\state(exp)$ or $\state(\varphi)$ we denote the number resp. Boolean value which
results from replacing every variable $v$ by the number $\state(v)$ in the arithmetic
expression $exp$ or the formula $\varphi$.
\begin{definition}[Closed Forms]\label{Closed Form Def}
	For a loop $(\guard, \update)$, an arithmetic expression $\normalfont{\cl{x_i}}$ is a \emph{closed form} for $x_i$ with \emph{start value} $n_0 \in \NN$ if $\normalfont{\cl{x_i}} = \sum_{1 \leq j \leq \ell} \alpha_j\cdot n^{a_j}\cdot b_j^n$ with      $\ell, a_j\in\NN$, $b_j\in\AA$,\footnote{$\AA$ is the set of algebraic numbers, i.e., the field of all roots of polynomials in $\ZZ[x]$.}  $\alpha_j\in\AA[\VSet]$,  and for all $\state:\VSet\cup \{n\}\to \ZZ$ with $\state(n) \geq n_0$ we have $\state(\normalfont{\cl{x_i}}) = \state(\update^n(x_i))$.
\pagebreak[2]	Similarly, we call $\cl{} = (\normalfont{\cl{x_1}},\ldots,\normalfont{\cl{x_d}})$
        a \emph{closed form} of the update $\update$ (resp.\ for the loop $(\guard,\update)$)
        with start value $n_0$ if for all $1 \leq i \leq d$, $\normalfont{\cl{x_i}}$ are
        closed forms for $x_i$ with start value $n_0$.
	\end{definition}

\begin{example}
	\label{ex:closed forms}
	In \cref{sect:solvable_loops} we will show that for the loop \eqref{WhileExample}, a closed form for $x_1$ (with start value 0) is $\normalfont{\cl{x_1}} = \tfrac{1}{2}\cdot\alpha\cdot (-\im)^n + \tfrac{1}{2}\cdot\compconj{\alpha}\cdot \im^n$ where $\alpha = (1 + 3\im)\cdot x_1 + 2\im\cdot x_2$.
	Here, $\compconj{\alpha}$ denotes the complex conjugate of $\alpha$, i.e., the sign of those monomials is flipped where the coefficient is a multiple of the imaginary unit $\im$.
	A closed form for $x_4$ (also with start value 0) is $\normalfont{\cl{x_4}} = x_4 + n \cdot (\tfrac{1}{6} + x_3 + x_3^2 - x_3\cdot n -\tfrac{n}{2}+\tfrac{n^2}{3})$.
\end{example}

Our
aim is to compute \emph{bounds} on the sizes of variables and on the runtime.
As in \cite{brockschmidt2016AnalyzingRuntimeSize,Festschrift}, we only consider bounds which are weakly monotonically increasing in all occurring variables.
Their advantage is that we can compose them easily (i.e., if $f$ and $g$ increase monotonically, then so does $f\circ g$).

\begin{definition}[Bounds]
	\label{def:bounds}
	The set of \emph{bounds}
	$\BoundSet$ is the smallest set with $\overline{\NN} = \NN \cup \{ \omega \} \subseteq \BoundSet$, $\VSet \subseteq \BoundSet$, and $\{b_1+b_2, \, b_1 \cdot b_2, \, k^{b_1}\} \subseteq \BoundSet \text{ for all } k \in \NN$ and $b_1,b_2 \in \BoundSet$.
\end{definition}

Size bounds should be bounds on the values of variables up to the point where the loop guard is not satisfied anymore for the first time.
To define size bounds,\linebreak
we introduce the \emph{runtime complexity} of a loop (whereas we considered the runtime complexity of arbitrary integer programs in \cite{brockschmidt2016AnalyzingRuntimeSize,Festschrift,lommen2022AutomaticComplexityAnalysis}).
Let $\State$ denote the set of all states $\state: \VSet \to \ZZ$ and let
$|\state|$ be the state with $|\state|(x) = |\state(x)|$ for all $x \in \VSet$.
\begin{definition}[Runtime Complexity for Loops]
	The \emph{runtime complexity} of\linebreak
	a loop $\IntLoop$ is $\rc : \State \rightarrow \NNC$ with $\rc(\state) = \inf\braced{n\in\NN\mid \state(\update^n( \neg\guard))}$, where $\inf \emptyset\linebreak
		= \omega$.
	An expression $r \in \BoundSet$ is a \emph{runtime bound} if $|\state|(r)\geq \rc(\state)$ for all $\state\in\State$.
\end{definition}
\begin{example}
	The runtime complexity of the loop \eqref{WhileExample} is $\rc(\state) = \max(0,\state(x_3))$.
	For example, $x_3$ is a runtime bound, as $|\state|(x_3) \geq \max(0,\state(x_3))$ for all states $\state\in\State$.
\end{example}

A \emph{size bound} on a variable $x$ is a bound on the absolute value of $x$ after $n$ iterations of the update $\update$, where $n$ is bounded by the runtime complexity.
In contrast to the definition of size bounds for transitions in integer programs from \cite{brockschmidt2016AnalyzingRuntimeSize}, \Cref{Size Bounds of Loops} requires that size bounds also hold \emph{before} evaluating the loop.
\begin{definition}[Size Bounds for Loops]
	\label{Size Bounds of Loops}
	$\Size: \VSet \rightarrow\BoundSet$ is a \emph{size bound} for $\IntLoop$ if for all $x\in\VSet$ and all $\state\in\State$, we have $|\state| (\Size(x)) \geq \sup\braced{|\state(\update^n(x))| \mid n \leq \rc(\state)}$.
\end{definition}

For any algebraic number $c \in \AA$, as usual $\ceil*{|c|}$ is the smallest natural number which is greater or equal to $c$'s absolute value.
Similarly, for any poly-exponen\-tial expression $p = \sum_{j} (\sum_i c_{i,j}\cdot
\beta_{i,j})\cdot n^{a_j} \cdot b_j^n$ where $c_{i,j} \in \AA$ and
the $\beta_{i,j}$ are normalized mono\-mials of the form $x_1^{e_1} \cdot \ldots \cdot x_d^{e_d}$,  $\ceil*{|p|}$ denotes $\sum_{j}
	\left(\sum_i \ceil*{|c_{i,j}|} \cdot \beta_{i,j}\right)\cdot n^{a_j} \cdot \ceil*{|b_j|}^n\!$.

We now determine size bounds by over-approximating the closed form $\cl{x}$ by the non-negative, weakly monotonically increasing function $\ceil*{|\cl{x}|}$.
Then we substitute $n$ by a runtime bound $r$ (denoted by ``$[n/r]$'').
Due to the monotonicity, this results in a bound on the size of $x$ not only at the end of the loop, but also during the iterations of the loop.
Since the closed form is only valid for $n$ iterations with $n \geq n_0$, we ensure that our size bound is also correct for less than $n_0$ iterations by symbolically evaluating the update, where we over-approximate maxima by sums.
As mentioned, see \cref{app:proofs} for the proofs of all new results.
\begin{theorem}[Size Bounds for Loops with Closed Forms]
	\label{thm:size_bounds_closed_form}
	Let \pagebreak[2] {\normalfont{$\cl{}$}} be a closed form  for the loop $\IntLoop$ with start
        value $n_0$ and let
	$r\in\BoundSet$ be a runtime bound.
	Then the (absolute) size of $x\in\VSet$ is bounded by	\mbox{\small $\normalfont{\size^x = \ceil*{|\cl{x}|} [n / r] + \sum_{0 \leq i < n_0} |\update^i(x)|}$}.
	Hence, the function $\Size$ with {\normalfont{$\Size(x) = \size^x$}} for all $x \in \VSet$ is a size bound for $\IntLoop$.
\end{theorem}
\makeproof{thm:size_bounds_closed_form}{
	\begin{myproof}
		We have to prove that
		\begin{equation}
			\label{eq:size_bounds_sup}
			|\state| (\size^x) \geq \sup\braced{|\state(\update^n(x))| \mid n \leq \rc(\state)}
		\end{equation}
		holds for all states $\state:\VSet \to \ZZ$ and all $x\in\VSet$.
		First, note that every summand of $|\state| (\size^x)$ is non-negative for all $\state:\VSet \to \ZZ$.
		Let $n \leq \rc(\state)$ for a $\state:\VSet \to \ZZ$.
		If $n < n_0$, then \eqref{eq:size_bounds_sup} clearly holds as the summand $|\update^n(x)|$ occurs in $\size^x$ and every summand is non-negative.
		Otherwise, consider $n \geq n_0$.
		For all $\state:\VSet \to \ZZ$, the function $$ f(n) = |\state| (\ceil*{|\cl{x}|})$$ increases weakly monotonically in $n$.
		Since $r$ is a bound on the runtime complexity, i.e., $\state(r) \geq \rc(\state)$, we can conclude that $|\state|(\size^x) \geq f(|\state|(r)) \geq f(\rc(\state)) \geq f(n) \geq |\state(\cl{x})| = |\state(\update^n(x))|$ holds.
		Here, the second and third inequation hold because $f$ is weakly monotonically increasing and $|\state|(r) \geq \rc(\state) \geq n$.
		The last inequation holds as $f(n)$ over-approximates $|\state(\cl{x})|$.
		So in total we have that $|\state|(\size^x) \geq |\state(\update^n(x))|$ holds for all $n \leq \rc(\state)$, which proves \eqref{eq:size_bounds_sup}.
	\end{myproof}
}

\begin{example}
	\label{exa:sizeboundLoop}
	As mentioned, for the loop \eqref{WhileExample}, a closed form for $x_1$ with start value $0$ is $\normalfont{\cl{x_1}} =\tfrac{1}{2}\cdot\alpha\cdot (-\im)^n + \tfrac{1}{2}\cdot\compconj{\alpha}\cdot \im^n$ where $\alpha = (1 + 3\im)\cdot x_1 + 2\im\cdot x_2$.
	Hence, $\ceil*{|\cl{x_1}|} = \ceil*{| \tfrac{1}{2}\cdot\alpha\cdot (-\im)^n + \tfrac{1}{2}\cdot\compconj{\alpha}\cdot \im^n |}
		= (\ceil*{| \tfrac{1+ 3\rm{i}}{2}|}\cdot x_1 + \ceil*{|\im|}\cdot x_2 )\cdot \ceil*{|-\rm{i}|}^n + (\ceil*{| \tfrac{1- 3\rm{i}}{2}|}\cdot x_1 + \ceil*{| -\rm{i}|}\cdot x_2 )\cdot \ceil*{|\rm{i}|}^n = 4\cdot x_1 + 2\cdot x_2$,	as $\ceil*{| \tfrac{1+ 3 \rm{i}}{2}|} = \ceil*{| \tfrac{1- 3\rm{i}}{2}|} = \ceil*{\tfrac{\sqrt{10}}{2}} = 2$ and $\ceil*{| \rm{i}|} =\linebreak
		\ceil*{| -\rm{i}|} = 1$.
	So our approach infers \emph{linear} size bounds for $x_1$ and $x_2$ (the similar computations for $x_2$ are omitted) while \cite{brockschmidt2016AnalyzingRuntimeSize} only infers exponential size bounds.

	As this over-approximation does not depend on $n$, it directly yields a size bound, i.e., $\size^{x_1} = \ceil*{|\cl{x_1}|}$.
	In contrast, in the over-approximation $\ceil*{|\cl{x_4}|} = x_4 + n\left(1 + x_3 + x_3^2 + x_3\cdot n + n + n^2\right)$, we have to replace $n$ by a runtime bound like $x_3$.
	Thus, we obtain the overall size bound $\size^{x_4} = x_4 + 3\cdot x_3^3 + 2\cdot x_3^2 + x_3$.
\end{example}

Although this section focused on closed forms which are poly-exponential ex\-pressions, our technique is applicable to all loops where we can compute over-\linebreak
approximating bounds for the closed form and the runtime complexity.
For exam\-ple, the update $\update(x) = x^2$ has the closed form $x^{(2^n)}$, but
it does not admit a poly-\linebreak
exponential closed form due to $x$'s super-exponential growth.
However, by instantiating $n$ by a runtime bound, we can still compute a size bound for this update.
The reason for focusing on poly-exponential expressions is that we can compute such a closed form for all so-called \emph{solvable loops} automatically, see \Cref{sect:solvable_loops}.

%% file: solvable_loops.tex
\section{Size and Runtime Bounds for Solvable Loops}
\label{sect:solvable_loops}

In this section, we present a class of loops where our technique of \Cref{thm:size_bounds_closed_form} is ``complete''.
The technique relies on the computation of suitable closed forms and of runtime bounds.
In \Cref{sec:Closed Forms for Solvable Loops}, we show that poly-exponential closed forms can be computed for all \emph{solvable loops}
\cite{solvable-maps,xuSymbolicTerminationAnalysis2013,kincaidClosedFormsNumerical2019,frohn2020TerminationPolynomialLoops,kovacsReasoningAlgebraicallyPSolvable2008,DBLP:conf/vmcai/HumenbergerJK18}.
Then we prove in \Cref{sec:Periodic Rational Solvable Loops}
that finite runtime bounds are computable for all terminating solvable loops with only periodic rational eigenvalues.

A loop $\IntLoop$ is \emph{solvable}  if $\update$ is a \emph{solvable update} (see \Cref{def:solvable} below for a formal definition), which
partitions $\VSet$ into blocks $\SSet_1,\ldots, \SSet_m$
(and loop guards $\guard$ are not relevant for closed forms).
Each block allows updates with \emph{cyclic dependencies} between its variables and \emph{non-linear} dependencies on variables in blocks with lower indices.

\vspace*{-.01cm}
\begin{definition}[Solvable Update \cite{solvable-maps,xuSymbolicTerminationAnalysis2013,kincaidClosedFormsNumerical2019,frohn2020TerminationPolynomialLoops,kovacsReasoningAlgebraicallyPSolvable2008,DBLP:conf/vmcai/HumenbergerJK18}]
	\label{def:solvable}
	An update $\update:\VSet \to \ZZ[\VSet]$
        is \emph{solvable}
	if there exists a partition $\SSet_1, \ldots, \SSet_m$ of $\braced{x_1, \ldots,
          x_d}$ such that for all $1 \leq i \leq m$ we have
        $\vec{\update}_{\SSet_i} = A_{\SSet_i} \cdot \vec{x}_{\SSet_i} +
        \vec{p}_{\SSet_i}$
for an $A_{\SSet_i}\in\ZZ^{|\SSet_i|\times |\SSet_i|}$ and a $\vec{p}_{\SSet_i}\in\ZZ[\bigcup_{j < i} \SSet_j]^{|\SSet_i|}$,
        where $\vec{\update}_{\SSet_i}$ is the vector of all
        $\update(x_j)$ and $\vec{x}_{\SSet_i}$ is the vector of all $x_j$ with $j\in\SSet_i$.
	The eigenvalues of a solvable loop are defined as the union of the eigenvalues of all matrices $A_{\SSet_i}$.
	The loop is \emph{homogeneous} if
   $\vec{p}_{\SSet_i} = \vec{0}$ for all $1 \leq i \leq m$.
\end{definition}

\begin{example}
	\label{exa:solvable}
	The loop \eqref{WhileExample} is an example for a solvable loop using the partition $\SSet_1 = \braced{x_1,x_2}$, $\SSet_2 = \braced{x_3}$, and $\SSet_3 = \braced{x_4}$.
\end{example}

The crucial \pagebreak[2] idea for our results in \Cref{sec:Closed Forms for Solvable Loops,sec:Periodic Rational Solvable Loops} is to reduce the problem of\linebreak
finding closed forms and runtime bounds from solvable loops to \emphit{triangular weakly non-linear} loops (\emph{twn-loops}) \cite{frohn2019TerminationTriangularInteger,frohn2020TerminationPolynomialLoops,hark2020PolynomialLoopsTermination}.
A \emph{twn-update} is a solvable update where each block $\SSet_j$ has cardinality one.
Thus, a twn-update is \emph{triangular}, i.e., the update of a variable does not depend on variables with higher indices.
Furthermore, the update is \emph{weakly non-linear}, i.e.,
a variable does not occur non-linear in its own update.
We are mainly interested in loops over $\ZZ$,
but to handle solvable updates, we will transform them into twn-updates with coefficients from $\AA$.

\begin{definition}[TWN-Update \cite{frohn2020TerminationPolynomialLoops,hark2020PolynomialLoopsTermination,frohn2019TerminationTriangularInteger}]
	\label{def:twn-loop}
	An update $\update: \VSet\rightarrow\AA[\VSet]$ is \emph{twn}
	if for all $1 \leq i \leq d$ we have $\update(x_i) = c_i\cdot x_i + p_i$ for some $c_i \in \AA$ and some polynomial $p_i\in\AA[x_1, \ldots,x_{i-1}]$.
	A loop with a twn-update is called a \emph{twn-loop}.
\end{definition}
\noindent
Clearly, \eqref{WhileExample} is not a twn-loop due to the cyclic dependency between $x_1$ and $x_2$.

\subsection{Closed Forms for Solvable Loops}
\label{sec:Closed Forms for Solvable Loops}

\Cref{lem:transform_solvable} (which extends \cite[Thm.\	16]{frohn2020TerminationPolynomialLoops}
from solvable updates with real eigenvalues to arbitrary solvable updates) illustrates that one can transform any \underline{s}olvable update $\update_s$ into a \underline{t}wn-update $\update_t$ by an automorphism $\vartheta$.
Here, $\vartheta$ is induced by the change-of-basis matrix of the Jordan normal form of each block of $\update_s$.
Note that the Jordan normal form is always computable in polynomial time (see \cite{cai1994ComputingJordanNormal}).
\begin{lemma}
	[Transforming Solvable Updates (see {\cite[Thm.\ 16]{frohn2020TerminationPolynomialLoops}})] \label{lem:transform_solvable}
	Let $\update_s$ be a solvable update.
	Then $\vartheta: \VSet\rightarrow\AA[\VSet]$ is an automorphism, where $\vartheta$ is defined by $\vartheta(\SSet) = P\cdot \xvec_\SSet$ for each block $\SSet$, where $J(A_{\SSet}) = P\cdot A_{\SSet}
		\cdot P^{-1}$ is the Jordan normal form of $A_\SSet$.
	Furthermore, $\update_t = \vartheta^{-1}\circ\update_s\circ \vartheta$ is a twn-update.
\end{lemma}
\begin{example}
	\label{exa:transform_solvable}
	To illustrate \Cref{lem:transform_solvable}, we transform the solvable update $\update_s$ of \eqref{WhileExample}
	into a twn-update $\update_t$.
	As the blocks $\SSet_2 = \braced{x_3}$ and $\SSet_3 = \braced{x_4}$ have cardinality one, we only have to consider $\SSet_1 = \braced{x_1,x_2}$.
	The restriction of $\update_s$ to $\SSet_1$ is $\myvec{x_1\\
			x_2} \leftarrow A_{\SSet_1} \cdot \myvec{x_1\\
			x_2}$ with $A_{\SSet_1} =
		\begin{sbmatrix}
			3 & 2 \\
			-5 & -3
		\end{sbmatrix}
	$.
	So we get the Jordan normal form $J(A_{\SSet_1}) = P\cdot A_{\SSet_1} \cdot P^{-1} =
		\begin{sbmatrix}
			-\!\im & 0 \\
			0 & \im
		\end{sbmatrix}
	$ where $P =
		\begin{sbmatrix}
			- \tfrac{5}{2}\im & \;\tfrac{1}{2}(1\!-\!3\im) \\
			\phantom{-} \;\, \tfrac{5}{2}\im & \;\tfrac{1}{2}(1\!+\!3\im)
		\end{sbmatrix}
	$ and $P^{-1} =
		\begin{sbmatrix}
			\tfrac{1}{5}(\im\!-\!3) & \;-\tfrac{1}{5}(\im\!+\!3) \\
			1 & \;1
		\end{sbmatrix}
	$. Thus, we have the following automorphism $\vartheta$ and its inverse $\vartheta^{-1}$:
	\[
		\mbox{\small $
				\begin{array}{l@{\quad}lll@{\qquad\quad}ll}
					                                                                               & \vartheta\myvec{x_1                                                      \\
					x_2}                                                                           & = P \cdot \myvec{x_1                                                     \\
					x_2}                                                                           & = \myvec{- \tfrac{5}{2}\im\cdot x_1 + \tfrac{1}{2}(1-3\im)\cdot x_2      \\
					\phantom{-} \;\,	 \tfrac{5}{2}\im\cdot x_1 + \tfrac{1}{2}(1 + 3\im)\cdot x_2 }, & \vartheta\myvec{x_3                                                      \\
					x_4}                                                                           & = \myvec{x_3                                                             \\
					x_4}                                                                                                                                                      \vspace*{0.1cm}\\
					                                                                               & \vartheta^{-1}\myvec{x_1                                                 \\
					x_2}                                                                           & = P^{-1} \cdot \myvec{x_1                                                \\
					x_2}                                                                           & = \myvec{\tfrac{1}{5}(\im - 3)\cdot x_1 - \tfrac{1}{5}(\im + 3)\cdot x_2 \\
					x_1 + x_2},                                                                    & \vartheta^{-1}\myvec{x_3                                                 \\
					x_4}                                                                           & = \myvec{x_3                                                             \\
						x_4}
				\end{array}
			$}
	\]
	\useshortskip Hence, $\update_t = \vartheta^{-1} \circ \update_s \circ \vartheta$ is the following twn-update:
	\useshortskip
	\[
		\update_t(x_1) = -\im\cdot x_1,\quad \update_t(x_2) = \im\cdot x_2,\quad \update_t(x_3) = x_3 - 1, \quad \update_t(x_4) = x_4 + x_3^2
	\]
\end{example}

The reason for transforming solvable updates to twn-updates is that for the latter, we can re-use our previous algorithm from \cite{frohn2019TerminationTriangularInteger} to compute poly-exponential closed forms.
While \cite{frohn2019TerminationTriangularInteger} only considered updates with linear arithmetic over $\ZZ$, it can directly be extended to twn-updates over $\AA$.

\begin{lemma}[Closed Forms for TWN-Updates (see \cite{frohn2019TerminationTriangularInteger})]
	\label{Closed Forms for TWN Updates}
	Let $\update$ be a twn-update.
	Then a (poly-exponential) closed form is computable for $\update$.
\end{lemma}

\begin{example}
	\label{exa:closed_form_twn}
	For $\update_t$ from \Cref{exa:transform_solvable}, we obtain the following closed form (with start value $0$): \pagebreak[2] $\clExp{}{t} = ((-\im)^n\cdot x_1, \im^n\cdot x_2, x_3 - n, x_4 + n(\tfrac{1}{6} + x_3 + x_3^2 - x_3\cdot n -\tfrac{n}{2}+\tfrac{n^2}{3}))$.
	\end{example}

So to obtain a closed form of a solvable update $\update_s$, we first transform it into a twn-update $\update_t$ via \Cref{lem:transform_solvable}, and then compute the closed form $\clExp{}{t}$ of $\update_t$ (\Cref{Closed Forms for TWN Updates}).
We now show how to obtain a closed form for $\update_s$ from $\clExp{}{t}$.
\begin{theorem}[Closed Forms for Solvable Updates]
	\label{thm:closed_form_solvable}
	Let $\update_s$ be a solvable update and $\vartheta$ be an automorphism as in \Cref{lem:transform_solvable}
	such that $\update_t = \vartheta^{-1}\circ\update_s\circ \vartheta$ is a twn-update.
	If $\normalfont{\clExp{}{t}}$ is a closed form of $\update_t$ with start value $n_0$, then $\normalfont{\clExp{}{s}} = \vartheta\circ\normalfont{\clExp{}{t}}\circ\vartheta^{-1}$ is a closed form of $\update_s$ with start value $n_0$.
\end{theorem}
\makeproof{thm:closed_form_solvable}{
	\begin{myproof}
		Let $\clExp{}{t}$ be a closed form for $\update_t$ with start value $n_0$.
		We have to prove that $\vartheta\circ\normalfont{\clExp{}{t}}\circ\vartheta^{-1}$ is a closed form for $\update_s$ with start value $n_0$.
		Let $\state: \VSet\rightarrow\ZZ$ be an arbitrary state and $x\in\VSet$.
		Note that $\update_t = \vartheta^{-1}\circ\update_s\circ \vartheta$ implies $\update_s = \vartheta\circ\update_t\circ \vartheta^{-1}$.

		Thus, for all $n \geq n_0$ we have
		\begin{align*}
			\state(\update_s^n(x)) & = \state((\update_s\circ\ldots\circ\update_s)(x))                                                                               \\
			                       & = \state(((\vartheta \circ \update_t \circ \vartheta^{-1})\circ\ldots\circ(\vartheta \circ \update_t \circ \vartheta^{-1}))(x)) \\
			                       & = \state((\vartheta \circ (\update_t \circ \ldots\circ \update_t) \circ \vartheta^{-1})(x))                                     \\
			                       & = \state((\vartheta \circ \update_t^n \circ \vartheta^{-1})(x))                                                                 \\
			                       & = \state((\vartheta \circ \clExp{}{t} \circ \vartheta^{-1})(x)).
		\end{align*}

                \hspace*{1cm}

                 \vspace*{-.8cm}
             	\end{myproof}
}

\begin{example}
	\label{exa:closed_form_solvable}
	In \cref{exa:transform_solvable} we transformed $\update_s$ into the twn-update $\update_t$ via an automorphism $\vartheta$ and in \cref{exa:closed_form_twn}, we gave a closed form $\normalfont{\clExp{}{t}}$ of $\update_t$.
	Thus, by \cref{thm:closed_form_solvable}, we can infer a closed form $\normalfont{\clExp{}{s}}= \vartheta\circ\normalfont{\clExp{}{t}}\circ\vartheta^{-1}$ of $\update_s$.
	For example, we compute a closed form for $x_1$ with start value $0$ ($\clExp{x_2}{s}$ can be inferred in a similar way):
	\begin{align*}
		\clExp{x_1}{s} & = \left(\tfrac{1}{5}(\im - 3)\cdot x_1 - \tfrac{1}{5}(\im +
                3)\cdot x_2\right)\; \left[v/\clExp{v}{t} \mid v\in\VSet \right] \; \left[v/\vartheta(v) \mid v\in\VSet \right]                                  \\
		               & = \left(\tfrac{1}{5}(\im - 3)\cdot(-\im)^n \cdot x_1 -
                \tfrac{1}{5}(\im + 3)\cdot\im^n \cdot x_2\right) \; \left[v/\vartheta(v) \mid v\in\VSet \right]                                                    \\
			               & = \tfrac{1}{2}(\underbrace{(1 + 3\im)\cdot x_1 + 2\im\cdot x_2}_{\alpha}) \cdot(-\im)^n + \tfrac{1}{2}(\underbrace{(1 - 3\im)\cdot x_1 - 2\im\cdot x_2}_{\compconj{\alpha}}) \cdot\im^n.
	\end{align*}
\end{example}

\subsection{Periodic Rational Solvable Loops}
\label{sec:Periodic Rational Solvable Loops}

In \Cref{sec:Closed Forms for Solvable Loops}, we discussed how to compute closed forms for solvable updates (by transforming them to twn-updates).
However to compute size bounds, we have to instantiate the variable $n$ in the closed forms by runtime bounds (\Cref{thm:size_bounds_closed_form}).
In \cite{hark2020PolynomialLoopsTermination}, it was shown that (polynomial) runtime bounds can always be computed for terminating twn-loops over the integers.
However, in general, transforming solvable loops via \Cref{lem:transform_solvable}
yields twn-updates which may contain algebraic (complex) numbers.
We now show that for the subclass of terminating \emph{periodic rational}
solvable loops, our approach is ``complete'' (i.e., finite runtime bounds and thus, also finite size bounds are always computable).

\begin{definition}[Periodic Rational \cite{kincaidClosedFormsNumerical2019}]
	A number $\lambda\in\AA$ is \emph{periodic rational} if $\lambda^p\in\QQ$ for some $p\in\NN$ with $p > 0$.
	The \emph{period} of $\lambda$ is the smallest such $p$ with $\lambda^p\in\QQ$.
	A solvable loop is \emph{periodic rational} (i.e., it is a \emph{prs loop}) with period $p$ if all its eigenvalues $\lambda$ are periodic rational and $p$ is the least common multiple of all their periods.
	A prs loop is a \emph{unit} prs loop if $|\lambda| \leq 1$ for all its eigenvalues $\lambda$.
\end{definition}
So $\im$, $-\im$, and $\sqrt{2} \cdot \im$ are periodic rational with period 2, while $\sqrt{2} + \im$ is not periodic rational.
The following lemma from \cite{kincaidClosedFormsNumerical2019}
gives a bound on the period of prs loops and thus yields an algorithm to detect prs loops and to compute their period.
\begin{lemma}[Bound on the Period \cite{kincaidClosedFormsNumerical2019}]
	\label{Bound on the Period}
	Let $A\in\ZZ^{n\times n}$.
	If $\lambda$ is a periodic rational eigenvalue of $A$ with period $p$, then $p \leq n^3$.
\end{lemma}

Now we show that by \emph{chaining} (i.e., by performing $p$ iterations of a prs loop with
period $p$ in a single step), one can transform any prs loop into a solvable loop with
only integer eigenvalues.
Then, our previous results on twn-loops
\cite{frohn2020TerminationPolynomialLoops,hark2020PolynomialLoopsTermination} can be used
to infer runtime bounds for these loops.

\begin{definition}[Chaining Loops]
	\label{def:chaining}
	Let $L = \IntLoop$ be a loop and  \pagebreak[2] $p \in \NN \setminus \{0\}$.
	Then $L_p = (\guard_p,\update_p)$ results from iterating $L$  $p$ times, i.e.,
	$\guard_p = \guard \, \land \, \update(\guard) \, \land \,
        \update(\update(\guard)) \, \land \, \ldots \,\, \land \, \update^{p-1}(\guard) \;
        \text{ and } \; \update_p(v) = \update^p(v) \text{ for all }
        v\in\VSet.$
	\end{definition}

\begin{example}
	\label{while example chained}
	The eigenvalues \mbox{\small $\pm\im$} of \eqref{WhileExample}
	have period 2.
	Chaining yields \mbox{\small $(\guard\!\wedge\!\update(\guard), \update^2)$}:
	\begin{equation}
		\label{WhileExampleChained}
		\mbox{\small $\textbf{while }\!(x_3\!>\!0 \wedge x_3\!>\!1)\!\textbf{ do }\! (x_1,x_2,x_3,x_4) \leftarrow (-x_1,-x_2,x_3\!-\!2,x_4\!+\!(x_3\!-\!1)^2\!+\!x_3^2)$}
	\end{equation}
\end{example}

Due to \cref{lem:transform_solvable} we can transform every solvable update into a twn-update by a (linear) automorphism $\vartheta$.
For prs loops, $\vartheta$'s range can be restricted to $\QQ[\VSet]$, i.e., one does not need algebraic numbers.
So, we first chain the prs loop $L$ and then compute a $\QQ$-automorphism $\vartheta$ transforming the chained loop $L_p$ into a twn-loop $L_t$ via \cref{lem:transform_solvable}.
Then we can infer a runtime bound for $L_t$ as in \cite{hark2020PolynomialLoopsTermination}.
The reason is that all factors $c_i$ in the update of $L_t$ are integers and thus,
we can compute a closed form $\sum_j \alpha_j\cdot n^{a_j}\cdot b_j^n$ such that $\alpha_j\in\QQ[\VSet]$ and $b_j\in\ZZ$.
Afterwards, the runtime bound for $L_t$ can be lifted to a runtime bound for the original loop by reconsidering the automorphism $\vartheta$.
Similarly, in order to prove termination of the prs loop $L$, we analyze termination of $L_t$ on $\vartheta(\ZZ^d) = \braced{\vartheta(\vec{x})\mid \vec{x}\in\ZZ^d}$.\footnote{By \cite{frohn2020TerminationPolynomialLoops}, termination of $L_t$ on $\vartheta(\ZZ^d)$ is reducible to invalidity of a formula $\exists \vec{x}\in\QQ^d .\linebreak
		\psi_{\vartheta(\ZZ^d)}\wedge \xi_{L_t}$. Here, $\psi_{\vartheta(\ZZ^d)}$ holds iff $\vec{x}\in\vartheta(\ZZ^d)$ and $\xi_{L_t}$ holds iff $L_t$ does not terminate\linebreak
	on $\vec{x}$.
	As shown in \cite{frohn2020TerminationPolynomialLoops}, non-termination of linear
        twn-loops with integer ei\-genvalues is \textsf{NP}-complete and it is semi-decidable for twn-loops with non-linear arithmetic.}

\begin{lemma}[Runtime Bounds for PRS Loops]
	\label{lem:correctness_chaining}
	Let $L$ be a prs loop with period $p$ and let $L_p = (\guard_p, \update_p)$ result from chaining as in \Cref{def:chaining}.
	From $\update_p$, one can compute a linear automorphism $\vartheta:\VSet\rightarrow\QQ[\VSet]$ as in \cref{lem:transform_solvable}, such that:
	\begin{enumerate}[(a)]
		\item $L_p$ is solvable and only has integer eigenvalues.
		\item $(\vartheta^{-1}\circ\update_p\circ\vartheta) : \VSet \to \QQ[\VSet]$ is a twn-update as in \Cref{def:twn-loop} such that all $c_i\in\ZZ$.
		\item $L_t = (\guard_t, \update_t)$ with $\guard_t = \vartheta^{-1}(\guard_p)$ and $\update_t = \vartheta^{-1}\circ\update_p\circ\vartheta$
		      is a twn-loop.
		      Moreover, the following holds:
					\vspace*{0.1cm}
					\begin{itemize}[label=\textbullet]
						\item $L$ terminates on $\ZZ^d$ iff
						\item $L_p$ terminates on $\ZZ^d$ iff
						\item $L_t$ terminates on $\vartheta(\ZZ^d) = \braced{\vartheta(\vec{x})\mid \vec{x}\in\ZZ^d}$.
					\end{itemize}
									\vspace*{0.1cm}
						\item If $r\!$ is a runtime bound\footnote{
			      More precisely, $|\state|(r)\geq \inf\braced{n\in\NN\mid \state(\update^n_t( \neg\guard_t))}$ must hold for all $\state : \VSet \to \vartheta(\ZZ^d)$.
		      } $\!\!\!$ for $L_t$, then \mbox{\small $p\!\cdot\!\ceil{|\vartheta(r)|}\!+\!p\!-\!1$} is a runtime bound for $L$.\linebreak
	\end{enumerate}
\end{lemma}
\makeproof{lem:correctness_chaining}{
	\begin{myproof}
		\begin{enumerate}[(a)]
			\item Consider a block $\SSet_i$ of $L$'s original update $\update$.
			      Then we have $\vec{\update}_{\SSet_i} = A_{\SSet_i} \cdot
                              \vec{x}_{\SSet_i} + \vec{p}_{\SSet_i}$ for an
                              $A_{\SSet_i}\in\ZZ^{|\SSet_i|\times |\SSet_i|}$ and a $\vec{p}_{\SSet_i}\in\ZZ[\bigcup_{j < i}
					      \SSet_j]^{|\SSet_i|}$, as in \Cref{def:solvable}.
			      For any $m \geq 1$, let $\vec{\update}^{\,m}(\vec{x}_{\SSet_i})$ denote the vector of $\update^m(x_j)$ for all $j \in \SSet_i$.
			      By induction on $m$, we show that $\vec{\update}^{\,m}(\vec{x}_{\SSet_i}) = A_{\SSet_i}^m \cdot \vec{x}_{\SSet_i} + \pvec{p}_{\!\SSet_i}'$ for some $\pvec{p}'_{\!\SSet_i}\in\ZZ[\bigcup_{j < i} \SSet_j]^{|\SSet_i|}$.
			      For $m = 1$ our claim trivially holds, because $\vec{\update}(\vec{x}_{\SSet_i}) = \vec{\update}_{\SSet_i}$.
			      For $m > 1$ we get
			      \begin{align*}
				      \vec{\update}^{\,m + 1}(\vec{x}_{\SSet_i}) = \update(\vec{\update}^{\,m}(\vec{x}_{\SSet_i})) & \stackrel{\text{IH}}{=}
				      \update( A_{\SSet_i}^m \cdot \vec{x}_{\SSet_i} + \pvec{p}_{\!\SSet_i}')                                                                                                                                                                                                           \\
				                                                                                                   & = A_{\SSet_i}^m \cdot \update(\vec{x}_{\SSet_i})+ \update(\pvec{p}_{\!\SSet_i}')                                                                                                   \\
				                                                                                                   & = A_{\SSet_i}^m \cdot ( A_{\SSet_i} \cdot \vec{x}_{\SSet_i} + \vec{p}_{\SSet_i}
				      ) + \update(\pvec{p}_{\!\SSet_i}')                                                                                                                                                                                                                                                \\
				                                                                                                   & = A^{m + 1}_{\SSet_i} \cdot \vec{x}_{\SSet_i} + \underbrace{A_{\SSet_i}^m \cdot \vec{p}_{\SSet_i} + \update(\pvec{p}_{\!\SSet_i}')}_{\in\ZZ[\bigcup_{j < i} \SSet_j]^{|\SSet_i|}}.
			      \end{align*}
			      Our claim implies that $\update_p$ still is solvable (with the same partition), since $\update_p(v) = \update^p(v)$ for all $v \in \VSet$ by the definition of chaining in \Cref{def:chaining}.

			      We now show that $A_{\SSet_i}^p$ only has integer eigenvalues.
			      The characteristic polynomial $\chi_{A_{\SSet_i}^p}$ of $A_{\SSet_i}^p$ is monic, since
			      \[
				      \chi_{A_{\SSet_i}^p} = \det(\lambda I - A_{\SSet_i}^p) = \sum_{\pi\in S}\sgn(\pi)\prod_{j = 1}^{|\SSet_i|}
				      (\lambda\delta_{j,\pi(j)}-(A_{\SSet_i}^p)_{j,\pi(j)})
			      \]
			      for the symmetric group $S$ of degree $|\SSet_i|$ and the
                              Kronecker delta function $\delta_{i,j}$, where
                              $\lambda^{|\SSet_i|}$ only results from the multiplication of the factors $(\lambda\delta_{j,\pi(j)}- \ldots)$ for $\pi = \text{id}$ and thus, $\sgn(\text{id}) = 1$.
			      Therefore, we have $q =1$ for every root $\frac{p}{q}\in\QQ$ of $\chi_{A_{\SSet_i}^p}$ by the rational root theorem (see e.g., \cite{bunt1988historical}).
			      The matrix $A_{\SSet_i}^p\in\ZZ^{|\SSet_i| \times |\SSet_i|}$ clearly only has eigenvalues in $\QQ$ by the definition of the period $p$.
			      Thus, every root of $\chi_{A_{\SSet_i}^p}$ and hence every eigenvalue is integer.
			\item When using the automorphism $\vartheta$ induced by the Jordan normal form, one obtains the twn-update $\vartheta^{-1}\circ\update_p\circ\vartheta$ by \cref{lem:transform_solvable}.
			      Here, all constants $c_i$ are in $\ZZ$ as they are the (integer) eigenvalues of $L_p$.

			      Now we show \pagebreak[2] that $\vartheta(v)\in\QQ[\VSet]$ for all $v\in\VSet$.
			      Let $\lambda\in\ZZ$ be an eigenvalue of
                              $A_{\SSet_i}\in\ZZ^{|\SSet_i|\times |\SSet_i|}$ for the
                              block $\SSet_i$ with geometric multiplicity $k$.
			      The change-of-basis matrix of the Jordan normal form
                              consists of columns $(A_{\SSet_i} - \lambda I)^{k-1}
                              \vec{u}$, \ldots, $(A_{\SSet_i} - \lambda I) \vec{u}$,
                              $\vec{u}$ for a suitable
                              $\vec{u}\in\ker((A_{\SSet_i} - \lambda
                              I)^k)$.
			      As $\vec{u}\in\ker((A_{\SSet_i} - \lambda I)^k)$ iff $(A_{\SSet_i} - \lambda I)^k\vec{u} = 0$, we have $\vec{u}\in\QQ^{|\SSet_i|}$ and thus all columns are also in $\QQ^{|\SSet_i|}$.
			\item $L_t$ is a twn-loop, since $\vartheta^{-1}\circ\update_p\circ\vartheta$ is a twn-update.
			      $L$ terminates on $\ZZ^d$ iff $L_p$ terminates on $\ZZ^d$ because chaining does not change the termination behavior \cite{frohn2019TerminationTriangularInteger}.
			      In \cite[Cor.\ 17]{frohn2020TerminationPolynomialLoops} it is shown that $L_p$ terminates on $\ZZ^d$ iff $L_t$ terminates on $\vartheta(\ZZ^d)$, i.e., the transformation via automorphisms also does not change the termination behavior.
			\item

			      For any state $\sigma: \VSet \to \ZZ$, we define the state $\vartheta(\state)$ by $(\vartheta(\state))(v) = \sigma(\vartheta(v))$ for all $v \in \VSet$.
			      Similar to \cite[Cor.\ 17]{frohn2020TerminationPolynomialLoops}, we now show that for any state $\state$ and any $n \in \NN$ we have $(\vartheta(\sigma)) (\update_t^n(\guard_t)) \Leftrightarrow \sigma(\update_p^n(\guard_p))$, i.e., a run of $L_p$ starting in the state $\sigma$ corresponds to a run of $L_t$ starting in the state $\vartheta(\sigma)$.
			      We have
			      \begin{align*}
				       & (\vartheta(\sigma)) (\update_t^n(\guard_t))                                                                       \\
				       & \Leftrightarrow (\vartheta(\sigma)) ((\vartheta^{-1}\circ\update_p\circ\vartheta)^n(\guard_t))                    \\
				       & \Leftrightarrow (\vartheta(\sigma)) ((\vartheta^{-1}\circ\update_p^n\circ\vartheta)(\guard_t))                    \\
				       & \Leftrightarrow (\vartheta(\sigma)) ((\vartheta^{-1}\circ\update_p^n\circ\vartheta\circ\vartheta^{-1})(\guard_p)) \\
				       & \Leftrightarrow (\vartheta(\sigma)) ((\vartheta^{-1}\circ\update_p^n)(\guard_p))                                  \\
				       & \Leftrightarrow \sigma ((\vartheta\circ\vartheta^{-1}\circ\update_p^n)(\guard_p))                                 \\
				       & \Leftrightarrow \sigma (\update_p^n(\guard_p))
			      \end{align*}
			      Thus, we have $\inf\braced{n\in\NN\mid \state(\update_p^n(\neg\guard_p))} = \inf\braced{n\in\NN\mid (\vartheta(\state))(\update_t^n(\neg\guard_t))}$ for all $\state \in \Sigma$.
			      Hence, if we have a runtime bound $r$ on the runtime complexity of $L_t$, i.e., $|\state|(r)\geq \inf\braced{n\in\NN\mid \state(\update_t^n(\neg \guard_t))}$ for all $\state : \VSet \to \vartheta(\ZZ^d)$, then applying $\vartheta$ to $r$ yields a runtime bound for $L_p$.
			      In other words, we have $\inf\braced{n\in\NN\mid \state(\update_p^n(\neg\guard_p))} = \inf\braced{n\in\NN\mid (\vartheta(\state))(\update_t^n(\neg\guard_t))} \leq |\vartheta(\state)|(r) = |\state(\vartheta(r))| \leq |\state|(\ceil*{|\vartheta(r)|})$ for all states $\state\in\State$.
			      Hence, $\ceil*{|\vartheta(r)|}$ is a runtime bound for $L_p$.

			      Similar as in \cite[Lemma 18]{hark2020PolynomialLoopsTermination}, we now show that
			      $\rc_L(\sigma) \leq p \cdot \rc_{L_p}(\sigma) +p -1$ holds for all $\sigma\in\State$, where $\rc_L$ and $\rc_{L_p}$ denote the runtime complexities of $L$ and $L_p$, respectively.
			      \[
				      \begin{array}{ll}
					                          & \rc_{L_p}(\sigma) = \inf\braced{n\in\NN\mid \state(\update_p^n( \neg\guard_p))}                                                                                                                            \\
					      \Longleftrightarrow & \forall n\!<\!\rc_{L_p}(\sigma). \; \state(\update_p^n(\guard_p) \land \neg \update_p^{\rc_{L_p}(\sigma)}(\guard_p))                                                                                       \\
					      \Longleftrightarrow & \forall n\!<\!\rc_{L_p}(\sigma). \; \state(\update^{p\cdot n}(\guard) \land \ldots \land \update^{p\cdot n + p -1}(\guard) \land                                                                           \\
					                          & \hspace*{2cm} \neg ( \update^{p\cdot \rc_{L_p}(\sigma)}(\guard_p) \land \ldots \land \update^{p\cdot \rc_{L_p}(\sigma) + p -1}(\guard_p) ) )                                                               \\
					      \Longleftrightarrow & \forall n\!<\!\rc_{L_p}(\sigma). \; \state(\update^{p\cdot n}(\guard) \land \ldots \land \update^{p\cdot n + p -1}(\guard) \land                                                                           \\
					                          & \hspace*{2cm}
					      ( \neg \update^{p\cdot \rc_{L_p}(\sigma)}(\guard_p) \lor \ldots \lor \neg \update^{p\cdot \rc_{L_p}(\sigma) + p -1}(\guard_p) ) )                                                                                                \\
					      \Longleftrightarrow & \forall n\!<\!p \cdot \rc_{L_p}(\sigma). \; \state(\update^{n}(\guard) \land ( \neg \update^{p\cdot \rc_{L_p}(\sigma)}(\guard_p) \lor ... \lor \neg \update^{p\cdot \rc_{L_p}(\sigma) + p -1}(\guard_p) )) \\
					      \Longleftrightarrow & p \cdot \rc_{L_p}(\sigma) \; \leq \; \rc_{L}(\sigma) \; \leq \; p\cdot \rc_{L_p}(\sigma) + p -1
				      \end{array}
			      \]

			      Thus, the runtime bound $\ceil*{|\vartheta(r)|}$ for $L_p$ yields the runtime bound $p\cdot\ceil*{|\vartheta(r)|} + p -1$ for $L$.
		\end{enumerate}
	\end{myproof}
}

\vspace*{-.4cm}

\begin{figure}[t]
	\centering
{\scriptsize	\begin{tikzpicture}[->,>=stealth',shorten >=1pt,auto,node distance=3.4cm,semithick,initial text=$ $]
		\node (r0) {\begin{tabular}{c}
				Runtime \\ Bound:
			\end{tabular}};
		\node (r1) [right of=r0,xshift=-1.4cm]{\begin{tabular}{c}
				prs loop $L$ \\
			$p\cdot\ceil{|\vartheta(r)|} + p - 1$
			\end{tabular}};
		\node (r2) [right of=r1,xshift=.6cm]{\begin{tabular}{c}
				$L_p$ \\
				$\ceil{|\vartheta(r)|}$
			\end{tabular}};
		\node (r3) [right of=r2,xshift=0.6cm]{\begin{tabular}{c}
				$L_t$ with 	$\update_t:\VSet\rightarrow\QQ[\VSet]$\\
				$r$ by \cite{hark2020PolynomialLoopsTermination,frohn2020TerminationPolynomialLoops}
			\end{tabular}};
		\node (s0) [below of=r0,yshift=1.6cm] {Size Bound:};
		\node (s1) [below of=r1,yshift=1.6cm] {\begin{tabular}{c}
				solvable loop $L$ \\
				$\clExp{}{s}$
			\end{tabular}};
		\node (s3) [below of=r3,yshift=1.6cm] {\begin{tabular}{c}
			$L_t'$ with	 	$\update_t':\VSet\rightarrow\AA[\VSet]$ \\
				$\clExp{}{t}$ by \cite{frohn2019TerminationTriangularInteger}
			\end{tabular}};
		\draw (r1) edge node {chaining} (r2);
		\draw (r1) edge [below,draw=none] node {\cref{lem:correctness_chaining} (a)} (r2);
		\draw (r2) edge node {\mbox{\scriptsize $\vartheta:\VSet\rightarrow\QQ[\VSet]$}} (r3);
		\draw (r2) edge [below,draw=none] node {\cref{lem:correctness_chaining} (b)} (r3);
		\node (h1) [below of=r1,yshift=3cm] {};
		\node (h2) [below of=r2,yshift=3cm] {};
		\node (h3) [below of=r3,yshift=3cm] {};
		\draw (h2) edge [below,bend left=12,decorate, decoration={snake,amplitude=1.5pt}] node {\cref{lem:correctness_chaining} (c) \& (d)} (h1.center);
		\draw (h3) edge [below,bend left=12,decorate, decoration={snake,amplitude=1.5pt}] node {\cref{lem:correctness_chaining} (c) \& (d)} (h2.center);
		\draw (r0) edge [->,right,loosely dashed] node
	{\cref{thm:size_bounds_closed_form}} (s0);
		\draw (s1) edge node {\cref{lem:transform_solvable} by $\vartheta':\VSet\rightarrow\AA[\VSet]$} (s3);
		\node (h1a) [below of=s1,yshift=3.05cm] {};
		\node (h3a) [below of=s3,yshift=3.05cm] {};
		\draw (h3a) edge [above,bend left=5,decorate, decoration={snake,amplitude=1.5pt}] node {\cref{thm:closed_form_solvable}} (h1a.center);
	\end{tikzpicture}}
	\caption{Illustration of Runtime and Size Bound Computations}
	\vspace*{-.2cm}
	\label{fig:illustration}
\end{figure}
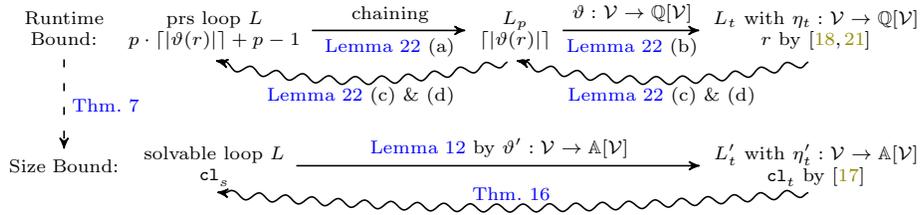

Since we can detect prs loops and their periods by \Cref{Bound on the Period},
\Cref{lem:correctness_chaining} allows us to compute runtime bounds for all terminating
prs loops. This is illustrated in
\Cref{fig:illustration}:
For runtime bounds, $L$ is transformed to $L_p$ by chaining and $L_p$ is transformed
further
to $L_t$ by an automorphism $\vartheta$.
The runtime bound $r$ for $L_t$ can then be transformed
into a runtime bound for $L_p$ and further into a runtime bound for $L$.  \pagebreak[2]
	For size bounds,  $L$ is directly transformed to a twn-loop $L_t'$  by an
        automorphism $\vartheta'$.
	The closed form 	$\clExp{}{t}$
        obtained for $L_t'$ is transformed via the automorphism $\vartheta'$ into a closed form	$\clExp{}{s}$ for $L$.
	Then the runtime bound for $L$ is inserted into this closed form to yield a size bound for $L$.
So in \Cref{fig:illustration}, standard arrows denote transformations of loops and wavy arrows denote
transformations of runtime bounds or closed forms.

\begin{theorem}[Completeness of Size and Runtime Bound Computation for Terminating PRS Loops]
	\label{thm:completeness}
	For all terminating prs loops, polynomial runtime bounds and finite	size bounds are computable.
	For terminating unit prs loops, all these size bounds are polynomial as well.

\end{theorem}
\makeproof{thm:completeness}{
	\begin{myproof}
		For a terminating prs loop $L$, we proceed as in \cref{lem:correctness_chaining}.
		So we first compute the loop $L_p$ via chaining (\cref{def:chaining}) which is still solvable (\cref{lem:correctness_chaining} (a)).
		Then we transform it via an automorphism $\vartheta$ into a twn-loop $L_t$ (\cref{lem:transform_solvable}) that is terminating on $\vartheta(\ZZ^d)$ (\cref{lem:correctness_chaining} (c)).
		As shown in \cite{hark2020PolynomialLoopsTermination}, since $\vartheta$ is a rational automorphism (\cref{lem:correctness_chaining} (b)), one can compute a polynomial runtime bound for $L_t$ on $\vartheta(\ZZ^d)$ and by \cref{lem:correctness_chaining} (d), this yields a polynomial runtime bound for $L$.

		By transforming the prs loop $L$ directly into a twn-loop as in \cref{lem:transform_solvable},
		from the closed form of the resulting twn-update we can obtain a closed form for $L$'s update (\Cref{thm:closed_form_solvable}).
		Thus, when inserting the runtime bound for $L$ into this closed form, \Cref{thm:size_bounds_closed_form} yields a finite size bound for $L$.

		Now we show that for terminating unit prs loops, we compute polynomial size bounds.
		Note that when transforming the prs loop $L$ into a twn-loop with update $\update_t'$, for all $1 \leq i \leq d$ we have $\update_t'(x_i) = c_i \cdot x_i + p_i$ where $c_i$ is an eigenvalue of $L$.
		The reason is that due to the definition of the transformation in \cref{lem:transform_solvable}, $c_i$ is on the diagonal of the Jordan normal form of $L$'s respective update matrix $A_\SSet$. Hence, we have $|c_i| \leq 1$, as $L$ is a unit prs loop. We call any twn-update of this form \emph{unit}.

		By induction on $i$, we now prove that any unit twn-update $\update_t'$ has closed forms $\cl{x_i} = \sum_{j = 1}^m \alpha_j \cdot n^{a_j} \cdot b_j^n$ where $|b_j| \leq 1$, $a_j\in\NN$, and $\alpha_i\in\AA[\VSet]$.
		We call such a poly-exponential expression \emph{unit}.
		\paragraph*{Induction Base:}
		We have $\update_t'(x_1) = c \cdot x_1 + p$ with $|c| \leq 1$ (as $\update_t'$ is unit) and $p\in\AA$.
		Moreover, we have $(\update_t')^n(x_1) = x_1 \cdot c^n + \sum_{i=1}^n
                c^{n-i}\cdot p = x_1 \cdot c^n + \frac{c^n - 1}{c - 1}\cdot p = (x_1 +
                \frac{p}{c - 1})\cdot c^n - \frac{p}{c - 1}$ if $c < 1$ and
                $(\update_t')^n(x_1) = x_1 + n\cdot p$ if $c = 1$.
		The last expressions are both unit closed forms $\cl{x_1}$.

		\paragraph*{Induction Step:}
		In the induction step we assume that we have unit closed forms $\cl{x_1},\ldots,\cl{x_k}$ for the variables $x_1,\ldots,x_k$.
		Similar to the induction base, we have $\update_t'(x_{k + 1}) = c \cdot x_{k + 1} + p$ with $|c| \leq 1$ and $p\in\AA[x_1,\ldots,x_k]$.
		So following \cite[Lemma 14]{frohn2019TerminationTriangularInteger}, we have to show that $(\update_t')^n(x_{k+1}) = c^n \cdot x_{k + 1} + \sum_{i = 1}^n c^{n-i} \cdot p[x_j / \cl{x_j} \mid j \in \braced{1,\ldots,k}][n / i - 1]$ can be transformed into a unit closed form.

		By the induction hypothesis, the closed forms $\cl{x_1},\ldots,\cl{x_k}$ are unit.
		We insert them into $p$ such that we obtain the expression $\sum_{j = 1}^m
                \alpha_j \cdot n^{a_j} \cdot b_j^n$ (see \eqref{thm:completeness_label1} below).
		This expression is unit, because being unit is closed under addition and multiplication, i.e., if $p,q$ are unit poly-exponential expressions, then $p + q$ and $p \cdot q$ are unit poly-exponential expressions as well.
		\begin{align}
			 & c^n \cdot x_{k + 1} + \sum_{i = 1}^n c^{n-i} \cdot p[x_j / \cl{x_j} \mid j \in \braced{1,\ldots,k}][n / i - 1] \nonumber                                                                                                                   \\
			 & = c^n \cdot x_{k + 1} + \sum_{i = 1}^n c^{n-i} \cdot \underbrace{\left(\sum_{j = 1}^m \alpha_j \cdot n^{a_j} \cdot b_j^n\right)}_{\coloneqq\; p[x_j / \cl{x_j} \mid j \in \braced{1,\ldots,k}]}[n / i - 1] \label{thm:completeness_label1}
			\\
			 & = c^n \cdot x_{k + 1} + \sum_{j = 1}^m \underbrace{\sum_{i = 1}^n c^{n - i} \cdot\alpha_j \cdot (i - 1)^{a_j} \cdot b_j^{i - 1}}_{\eqqcolon\; q_j} \label{thm:completeness_label2}
		\end{align}

		So, after commuting both sums in \eqref{thm:completeness_label2}, it remains to show that $q_j$ can be transformed into a unit poly-exponential expression.
		In \cite[Thm.\ 12]{frohn2019TerminationTriangularInteger} it is shown how to transform $q_j$ into a poly-exponential expression.
		Note that only $c^{n - i}$ or $b_j^{i - 1}$ lead to exponential factors $c^n$ resp.\ $b^n$.
		However, we have both $|b| \leq 1$ (by the induction hypothesis) and $|c| \leq 1$ (since $\update_t'$ is unit).

		To summarize, we have proven that \cref{Closed Forms for TWN Updates}
		yields unit closed forms $\sum_{j = 1}^m \alpha_j \cdot n^{a_j} \cdot b_j^n$ for the twn-loop resulting from transforming $L$.
		Remember that we obtain closed forms for solvable loops by reverting the automorphism (see \Cref{lem:transform_solvable}).
		This automorphism only affects the polynomials $\alpha_j \in \AA[\VSet]$, but it does not modify the exponential factors $b_j^n$.
		Thus, the closed forms for the original prs loop $L$ have the form $\cl{x} = \sum_{j = 1}^m \alpha_j' \cdot n^{a_j} \cdot b_j^n$ with $|b_j| \leq 1$, i.e., they are unit.
		Hence, we have $\ceil*{|\cl{x}|} = \sum_{j = 1}^m \ceil*{|\alpha_j'|}
			\cdot n^{a_j}
			\cdot \ceil*{|b_j|}^n = \sum_{j = 1}^{m} \ceil*{|\alpha_j'|} \cdot n^{a_j}$.
		Applying \Cref{thm:size_bounds_closed_form}, i.e., replacing $n$ by the polynomial runtime bound, finally yields a polynomial size bound.
	\end{myproof}
}

\begin{example}
	For the loop $L$ from \eqref{WhileExample}, we computed $L_p$ for $p=2$ in \eqref{WhileExampleChained}, see \Cref{while example chained}.\linebreak
As $L_p$ is already a twn-loop,
	we can use the technique of \cite{hark2020PolynomialLoopsTermination} (implemented in our\linebreak tool \KoAT{}) to obtain the runtime bound $x_3$ for $L_p$.
	\Cref{lem:correctness_chaining} yields the runtime bound
 $2\cdot x_3 + 1$  for the original loop \eqref{WhileExample}.
	Of course, here
	one could also use (incom\-plete) approaches based on linear ranking functions (also
        implemented in \KoAT{},\linebreak see, e.g.,
        \cite{brockschmidt2016AnalyzingRuntimeSize,Festschrift}) to directly infer the
        tighter runtime bound $x_3$ for the loop \eqref{WhileExample}.
\end{example}

%% file: integer_programs.tex
\section{Size Bounds for Integer Programs}
\label{sect:integer_programs}
Up to now, we focused on \emph{isolated} loops.
In the following, we incorporate our complete approach from \Cref{sect:size_bounds_closed_forms,sect:solvable_loops} into the setting of general \emph{integer programs} where most questions regarding termination or complexity are undecidable.
Formally, an integer program is a tuple $\IntProgram$ with a finite set of variables $\VSet$, a finite set of locations $\LSet$, a fixed initial location $\location_0 \in \LSet$, and a finite set of transitions $\TSet$.
A \emph{transition} is a 4-tuple $(\location,\guard,\update,\location')$ with a \emph{start location} $\location\in\LSet$, \emph{target location} $\location'\in\LSet\setminus\braced{\location_0}$, \emph{guard} $\guard\in\FormulaSet(\VSet)$, and \emph{update} $\update: \VSet\rightarrow\ZZ[\VSet]$.
To simplify the presentation, we do not consider ``temporary'' variables (whose update is non-deterministic), but the approach can easily be extended accordingly.
Transitions $(\location_0,\_,\_,\_)$ are called \emph{initial} and
$\TSet_0$ denotes the set of all initial transitions.

\begin{example}
	In the integer program of \cref{fig:ITS_solvable_loop}, we omitted identity updates $\update(v) =\linebreak
		v$ and guards where $\guard$ is $\true$.
	Here, $\VSet = \braced{x_1,\ldots,x_5}$ and $\LSet = \{\location_0, \location_1, \location_2\}$, where $\location_0$ is the initial location.
	Note that the loop in \eqref{WhileExample} \emph{corresponds} to transition $t_1$.
	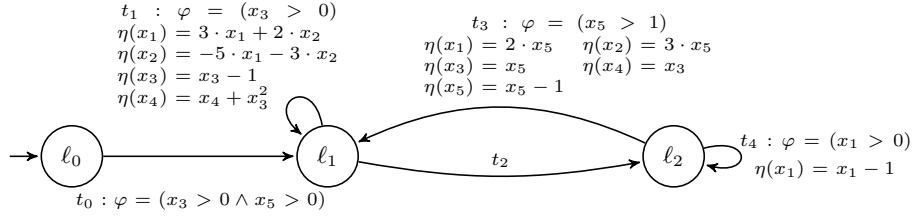
\begin{figure}[t]
		\centering
		\begin{tikzpicture}[->,>=stealth',shorten >=1pt,auto,node distance=3.5cm,semithick,initial text=$ $]
			\node[state,initial] (q0) {$\location_0$};
			\node[state] (q1) [right of=q0,xshift=-.1cm]{$\location_1$};
			\node[state] (q2) [right of=q1, node distance=4cm,xshift=.6cm]{$\location_2$};
			\draw (q0) edge node [text width=3.5cm,align=center,below,yshift=-.35cm]
				{\scriptsize $t_0: \guard = (x_3 > 0 \wedge x_5 > 0)$} (q1);
			\draw (q2) edge[bend left=-25] node [text width=4cm,align=center,above,xshift=0.9cm]
				{{\scriptsize {{\scriptsize $t_3 : \guard = (x_5 > 1)$\\
										$
											\begin{array}{rclcrl}
												\update(x_1) & = & 2 \cdot x_5 \quad & \update(x_2) & = & 3 \cdot x_5 \\
												\update(x_3) & = & x_5 \quad         & \update(x_4) & = & x_3        \\
												\update(x_5) & = & x_5 - 1           &              &   &             \\
											\end{array}
										$}}}} (q1);
			\draw (q1) edge[out=100,in=140, min distance=0.7cm] node [text width=5cm,align=center,yshift=1.4cm,xshift=-3.5cm]
				{{\scriptsize $t_1:\guard = (x_3 > 0)$ \\
							$
								\begin{array}{rcl}
									\update(x_1) & = & 3\cdot x_1 + 2\cdot x_2  \\
									\update(x_2) & = & -5\cdot x_1 - 3\cdot x_2 \\
									\update(x_3) & = & x_3 - 1                  \\
									\update(x_4) & = & x_4 + x_3^2              \\
								\end{array}
							$}} (q1);
			\draw (q2) edge [loop right] node [text width=2.8cm,align=center,xshift=-.4cm]
				{{\scriptsize $t_4:\guard = (x_1 > 0)$ \\
							$
								\begin{array}{rcl}
									\update(x_1) & = & x_1 - 1 \\
								\end{array}
							$}} (q2);
			\draw (q1) edge[bend left=-10] node [text width=3cm, above, align=center]
				{{\scriptsize $t_2$}} (q2);
		\end{tikzpicture} \vspace*{-.5cm}
		\caption{An Integer Program with Non-Linear Size Bounds}\label{fig:ITS_solvable_loop}
		\vspace*{-.4cm}
	\end{figure}
\end{example}

\begin{definition}[Correspondence between Loops and Transitions]
	\label{Correspondence between Loops and Transitions}
	Let $t = (\location, \guard, \update,\location)$ be a transition with $\guard\in\FormulaSet(\VSet')$ for some variables $\VSet'\subseteq\VSet$ such that $\update(x) \in \ZZ[\VSet']$ for all $x \in \VSet'$.
	A loop $(\guard', \update')$ with $\guard'\in\FormulaSet(\braced{x_1,\ldots,x_d})$
        and $\update': \braced{x_1,\ldots,x_d}\rightarrow\ZZ[\braced{x_1,\ldots,x_d}]$
        \emph{corresponds} to the transition $t$ via the variable renaming $\pi:
        \braced{x_1,\ldots,x_d}\rightarrow \VSet'$ if $\guard$ is $\pi(\guard')$ and for
        all $1\leq i \leq d$ we have  \pagebreak[2] $\update(\pi(x_i)) = \pi(\update'(x_i))$.
\end{definition}

To define the semantics of integer programs,
an evaluation step moves from one configuration $(\location,\state)\in\LSet\times\State$ to another configuration $(\location',\state')$ via a transition $(\location, \guard, \update, \location')$ where $\state(\guard)$ holds.
Here, $\state'$ is obtained by applying the update $\update$ on $\state$.
From now on, we fix an integer program $\PP = \IntProgram$.

\begin{definition}
	[Evaluation of Programs] For configurations $(\location,\state)$, $(\location',\state')$ and $t = (\location_t,\guard,\update,\location_{t}')\in\TSet$, $(\location,\state)\rightarrow_t(\location',\state')$ is an	\emph{evaluation}	step if $\location = \location_t$, $\location' = \location_{t}'$, $\state(\guard) = \normalfont{\true}$, and $\state(\update(v)) = \state'(v)$ for all $v\in\VSet$.
	Let $\to_{\TSet} \; = \, \bigcup_{t \in \TSet} \to_t$, where we also write $\to$ instead of $\to_t$ or $\to_{\TSet}$.
	Let $(\location_0,\state_0)\rightarrow^k(\location_k,\state_k)$ abbreviate $(\location_0,\state_0)\linebreak
		\rightarrow \ldots \rightarrow(\location_k,\state_k)$ and let $(\location,\state) \rightarrow^*(\location',\state')$ if $(\location,\state) \rightarrow^k(\location',\state')$ for some $k \geq 0$.
\end{definition}

\begin{example}
	If we encode states as tuples $(\state(x_1),\ldots,\state(x_5))\in\ZZ^5$, then $(-6,
		-8,\linebreak 2,1,1) \rightarrow_{t_0} (-6,-8,2,1,1) \rightarrow^2_{t_1} (6,8,0,6,1) \rightarrow_{t_2} (6,8,0,6,1)\rightarrow_{t_4}^6 (0,8,0,6,1)$.
\end{example}

Now we define size bounds for variables $v$ after evaluating a transition $t$:
$\Size(t,v)$ is a \emph{size bound} for $v$ w.r.t.\ $t$ if for any run starting in $\initial\in\State$, $|\initial|(\Size(t, v))$ is greater or equal to the largest absolute value of $v$ after evaluating $t$.
\begin{definition}[Size Bounds \cite{brockschmidt2016AnalyzingRuntimeSize,Festschrift}]
	\label{sizebounds}
	A function $\Size: (\TSet \times \VSet) \rightarrow \BoundSet$ is a \emph{(global) size bound} for the program $\PP$ if for all $(t, x) \in \TSet \times \VSet$ and all states $\initial \in \State$ we have $|\initial|(\Size(t, x)) \geq \sup\sizeboundtermx$.
\end{definition}

Later in \Cref{lem:lift_size_bounds_loops}, we will compare the notion of size bounds for transitions in a program from \Cref{sizebounds} to our earlier notion of size bounds for loops from \cref{Size Bounds of Loops}.

\begin{example}
	\label{ex:size bounds t0 t3}
	As an example, we give size bounds for the transitions $t_0$ and $t_3$ in \cref{fig:ITS_solvable_loop}.
	Since $t_0$ does not change any variables, a size bound is $\Size(t_0, x_i) = x_i$ for all $1 \leq i \leq 5$.
	Note that the value of $x_5$ is never increased and is bounded from below by $0$ in any run through the program.
	Thus, $\Size(t_3, x_3) = x_5 = \Size(t_3, x_5)$.
	Similarly, we have $\Size(t_3, x_1) = 2\cdot x_5$, $\Size(t_3, x_2) = 3\cdot x_5$, and $\Size(t_3, x_4) = x_3$.
	\end{example}

To infer size bounds for transitions as in \Cref{sizebounds} automatically, we lift \emph{local} size bounds (i.e., size bounds which only hold for a subprogram with transitions $\TSet'\subseteq\TSet\setminus\TSet_0$) to global size bounds for the \emph{complete}
program.
For the subprogram, one considers runs which start after evaluating an \emph{entry transition} of $\TSet'$.

\begin{definition}[Entry Transitions \cite{brockschmidt2016AnalyzingRuntimeSize}]
	\label{Entry Transitions}
	Let $\emptyset\neq\TSet' \subseteq \TSet\setminus\TSet_0$.
	The \emph{entry transitions} of $\TSet'$ are $ \entry_{\TSet'} = \braced{t \mid t\!=\!(\_,\_,\_,\location)\!\in\!\TSet\setminus\TSet' \text{ and there is a } (\location,\_,\_,\_)\!\in\!\TSet'}$.
\end{definition}

\begin{example}
	\label{ex:entry transitions}
	For the program in \cref{fig:ITS_solvable_loop},
	we have $\entry_{\braced{t_1}} = \{t_0, t_3\}$ and $\entry_{\braced{t_4}} = \braced{t_2}$.
\end{example}

\begin{definition}[Local Size Bounds]
	\label{def:local_size_bounds}
	Let $\emptyset\neq\TSet'\subseteq\TSet\setminus\TSet_0$ and $t'\in\TSet'$.
	$\Size_{t'}:\VSet\rightarrow\BoundSet$ is a \emph{local size bound} for $t'$ w.r.t.\ $\TSet'$ if for all $x\in\VSet$ and all $\state\in\State$:\footnote{To simplify the formalism, in this definition, we consider every possible configuration $(\location,\state)$ and not only configurations which are reachable from the initial location $\location_0$.}
	$|\state| (\Size_{t'}(x)) \geq \sup\braced{|\state'(x)|\mid \exists \location' \in \LSet, (\_,\_,\_,\location)\in\entry_{\TSet'}.\; (\location,\state)\; (\rightarrow^*_{\TSet'}\circ \rightarrow_{t'}) \; (\location',\state')}$.
\end{definition}

\Cref{thm:lift_size_bounds} below yields a novel \emph{modular} procedure to infer (global) size bounds from previously computed local size bounds.
A local size bound for a transition $t'$ w.r.t.\ a subprogram $\TSet'\subseteq\TSet\setminus\TSet_0$ is lifted by inserting size bounds for all entry transitions.
 Again, this is possible because we only use weakly monotonically increasing functions as bounds.
Here, ``$b\left[v/p_v \mid v\in\VSet \right]$'' \pagebreak[2] denotes the bound which results from replacing every variable $v$ by $p_v$ in the bound $b$.
\begin{theorem}[Lifting Local Size Bounds]
	\label{thm:lift_size_bounds}
	Let $\emptyset\neq\TSet'\subseteq\TSet\setminus\TSet_0$, let $\Size_{t'}$ be a local size bound for a transition $t'$ w.r.t.\ $\TSet'$ and let $\Size: (\TSet\times\VSet)\rightarrow\BoundSet$ be a size bound for $\PP$.
	Let $\Size'(t',x) = 	\sum_{\pret \in \entry_{\TSet'}} \Size_{t'}(x) \left[v/\Size(\pret,v) \mid v\in\VSet \right]$ and 	$\Size'(t,x) = \Size(t,x)$ for all $t' \neq t$.
	Then $\Size'$ is also a size bound for $\PP$.
	\end{theorem}
\makeproof{thm:lift_size_bounds}{
	\begin{myproof}
		We show that for all $(t,x)\in\TSet\times\VSet$ and all $\initial\in\State$ we have $$|\initial|(\Size'(t,x)) \geq \sup\sizeboundterm.$$

			The case $t' \neq t$ is trivial, as we have $\Size'(t,x) = \Size(t,x)$ for all $x\in\VSet$ and $\Size$ is a size bound for $\PP$.
		Otherwise, any evaluation with $\rightarrow^* \circ \rightarrow_{t'}$ starting in an initial configuration $(\location_0, \state_0)$ has the following form: $$(\location_0,\state_0)\, \rightarrow^* \circ \rightarrow_{\pret'}\, (\location,\state)\,\rightarrow_{\TSet'}^* \circ\rightarrow_{t'} \,(\location', \state') \quad \text{for some }
			\pret'\in\entry_{\TSet'}.$$

		Thus, we have to prove for all such evaluations and all variables $x\in\VSet$ that $|\state_0|(\Size'(t',x)) \geq |\state'(x)|$ holds:
		\[
			\begin{array}{rcl}
				|\state_0|(\Size'(t',x)) & =    & |\state_0|\left(\sum_{\pret \in \entry_{\TSet'}} \Size_{t'}(x) \left[v/\Size(\pret,v) \mid v\in\VSet \right]\right)                                                                                                  \\
				                         & \geq & |\state_0|\left(\Size_{t'}(x) \left[v/\Size(\pret',v) \mid v\in\VSet \right]\right)                                                                                                                                  \\
				                         &      & \qquad\text{(as every summand is non-negative)}                                                                                                                                                                      \\
				                         & \geq & \Size_{t'}(x) \left[v/|\state(v)| \mid v\in\VSet \right]                                                                                                                                                             \\
				                         &      & \qquad\parbox{8cm}{(as $|\state_0|(\Size(\pret', v)) \geq |\state(v)|$ for all $v\in\VSet$ since $\Size$ is a size bound for $\PP$ and $\Size_{t'}(x)$ is a weakly monotonically increasing bound from $\BoundSet$)} \\
				                         & =    & |\state|(\Size_{t'}(x))                                                                                                                                                                                              \\
				                         & \geq & |\state'(x)|                                                                                                                                                                                                         \\
				                         &      & \qquad \parbox{8cm}{(by \Cref{def:local_size_bounds}, since $\Size_{t'}$ is a local size bound for transition $t'$)}
			\end{array}
		\]
	\end{myproof}
}

To obtain local size bounds which can then be lifted via \Cref{thm:lift_size_bounds}, we look for transitions $t_L$ that correspond to a loop $L$ and then we compute a size bound for $L$ as in \Cref{sect:size_bounds_closed_forms,sect:solvable_loops}.
The following lemma shows that size bounds for loops as in \Cref{Size Bounds of Loops}
indeed yield local size bounds for the corresponding transitions.\footnote{\label{size
  bounds loops n 0}Local or global size bounds for transitions only have to hold if the
transition is indeed taken.
	In contrast, size bounds for loops also have to hold if there is no loop iteration.
	This will be needed in \Cref{thm:cycles} to compute local size bounds for simple cycles.}

\begin{lemma}[Local Size Bounds via Loops]
	\label{lem:lift_size_bounds_loops}
	Let	$\Size_L$ be a size bound for a loop $L$ (as in \cref{Size Bounds of Loops}) which corresponds to a transition $t_L$ via a variable renaming $\pi$.
	Then $\pi \circ \Size_L \circ \pi^{-1}$ is a local size bound for $t_L$ w.r.t.\ $\{t_L\}$ (as in \cref{def:local_size_bounds}).
\end{lemma}
\makeproof{lem:lift_size_bounds_loops}{
	\begin{myproof}
		Let $L = (\guard',\update')$ be a loop which corresponds to the transition $t = (\location, \guard, \update, \location)$ via the variable renaming $\pi$.
		By \cref{def:local_size_bounds}, we have to show that for all $x\in\VSet$ and all $\state\in\State$ we have
		\[
			\mbox{\small $ |\state|(\pi(\Size_L(\pi^{-1}(x)))) \geq \sup\{ |\state'(x)|\mid \exists \,(\_,\_,\_,\location)\in\entry_{\braced{t_L}}.\; (\location,\state) \, (\rightarrow^*_{t_L}\circ \rightarrow_{t_L}) \, (\location,\state')\}.$}
		\]

		Recall that $\rc: \State\rightarrow\overline{\NN}$ denotes the runtime complexity of $L$.
		Thus, we have
		\begin{align*}
			\;      & |\state|(\pi(\Size_L(\pi^{-1}(x))))                                                                                                                                                                 \\
			= \;    & |\state \circ \pi|(\Size_L(\pi^{-1}(x)))                                                                                                                                                            \\
			\geq \; & \sup\braced{|\state(\pi(\update'^n(\pi^{-1}(x))))| \mid n \leq \rc(\state \circ \pi)}
			\tag*{(by \cref{Size Bounds of Loops})}                                                                                                                                                                       \\
			= \;    & \sup\braced{|\state(\update^n(\pi(\pi^{-1}(x))))| \mid n \leq \rc(\state \circ \pi)}
			\tag*{(by the correspondence between $L$ and $t_L$)}                                                                                                                                                          \\
			= \;    & \sup\braced{|\state(\update^n(x))| \mid n \leq \rc(\state \circ \pi)}                                                                                                                               \\
			=	\;     & \sup\{ |\state'(x)|\mid \forall\, 0 \leq n \leq \rc(\state\circ \pi). \;(\location,\state)\rightarrow^n_{t_L} (\location,\state')\}                                                                 \\
			\geq	\;   & \sup\{ |\state'(x)|\mid \forall\, 1 \leq n \leq \rc(\state\circ \pi). \;(\location,\state)\rightarrow^n_{t_L} (\location,\state')\} \tag*{(by omitting $n = 0$ in the supremum)}                    \\
			= \;    & \sup\{ |\state'(x)|\mid (\location,\state) \; (\rightarrow^*_{t_L}\circ \rightarrow_{t_L}) \; (\location,\state')\}\tag*{(by the definition of $\rc$ and the correspondence between $L$ and $t_L$)} \\
			\geq \; & \sup\{ |\state'(x)|\mid \exists \, (\_,\_,\_,\location)\in\entry_{\braced{t_L}}.\; (\location,\state) \; (\rightarrow^*_{t_L}\circ \rightarrow_{t_L}) \; (\location,\state')\}
		\end{align*}
	\end{myproof}
}

\begin{example}
	\label{size bound t1}
	$\Size_L(x_4) = x_4 + 3\cdot x_3^3 + 2\cdot x_3^2 + x_3$ is a size bound for $x_4$ in the loop \eqref{WhileExample}, see \cref{exa:sizeboundLoop}.
		This loop corresponds to transition $t_1$ in the program of \cref{fig:ITS_solvable_loop}.
	Since $\entry_{\braced{t_1}} = \{t_0, t_3\}$ by \Cref{ex:entry transitions}, \cref{thm:lift_size_bounds} yields the following (non-linear) size bound for $x_4$ in the full program of \cref{fig:ITS_solvable_loop} (see \cref{ex:size bounds t0 t3} for $\Size(t_0,v)$ and $\Size(t_3,v)$):
	\begin{align*}
		\Size(t_1,x_4) & = \Size_L(x_4) \left[v/\Size(t_0,v) \mid v\in\VSet \right] + \Size_L(x_4) \left[v/\Size(t_3,v) \mid v\in\VSet \right] \\
		               & = (x_4 + 3\cdot x_3^3 + 2\cdot x_3^2 + x_3) + (x_3 + 3\cdot x_5^3 + 2\cdot x_5^2 + x_5)                               \\
		               & = 2\cdot x_3 + 2\cdot x_3^2 + 3\cdot x_3^3 + x_4 + x_5 + 2\cdot x_5^2 + 3\cdot x_5^3
	\end{align*}
	Analogously, we infer the remaining size bounds $\Size(t_1,x_i)$, e.g., $\Size(t_1,x_1)\!=\!(4\cdot x_1\linebreak
		+ 2\cdot x_2) \left[v/\Size(t_0,v) \mid v\!\in\!\VSet \right] + (4\cdot x_1 +\!
		2\cdot x_2)\left[v/\Size(t_3,v) \mid v\!\in\!\VSet \right] = 4\cdot x_1
		+
		2\cdot x_2 +
		14\cdot x_5$.
\end{example}

Our approach alternates between improving size and runtime bounds for indi\-vidual transitions.
We start with $\Size(t_0,x) = |\update(x)|$ for initial transitions $t_0\in\TSet_0$\linebreak
where $\update$ is $t_0$'s update, and $\Size(t,\_) = \omega$ for $t\in\TSet\setminus\TSet_0$.
Here, similar to the notion $\ceil*{|p|}$ in \Cref{sect:size_bounds_closed_forms}, for every polynomial $p = \sum_j c_{j}\cdot \beta_{j}$ with normalized monomials $\beta_j$, $|p|$ is the polynomial $\sum_j |c_{j}|\cdot \beta_{j}$.
To improve the size bounds\linebreak
of transitions that correspond to (possibly non-linear) solvable loops, we can use\linebreak
closed forms (\cref{thm:size_bounds_closed_form}) and the lifting via \cref{thm:lift_size_bounds}.
Otherwise, we use an existing incomplete technique \cite{brockschmidt2016AnalyzingRuntimeSize} to improve size bounds (where \cite{brockschmidt2016AnalyzingRuntimeSize}
essentially only succeeds for updates without non-linear arithmetic).
In this way, we can automatically compute polynomial size bounds for all remaining transitions and variables in the program of \cref{fig:ITS_solvable_loop} (e.g., we obtain $\Size(t_2,x_1) = \Size(t_1,x_1) = 4\cdot x_1 + 2\cdot x_2 + 14\cdot x_5$).

Both the technique from \cite{brockschmidt2016AnalyzingRuntimeSize}
and our approach from \cref{thm:size_bounds_closed_form}
rely on runtime bounds to compute size bounds.
On the other hand, as shown in \cite{brockschmidt2016AnalyzingRuntimeSize,Festschrift,lommen2022AutomaticComplexityAnalysis}, size bounds for ``previous'' transitions are needed to infer (global) runtime bounds for transitions in a program.
For that reason, the alternated computation resp.\ improvement of global size and runtime bounds for the transitions is repeated until all bounds are finite.
We will illustrate this in more detail in \Cref{sect:complexity}.

In \Cref{Correspondence between Loops and Transitions,lem:lift_size_bounds_loops} we
considered   \pagebreak[2] transitions with the same start and target location that directly correspond to loops.
To increase the applicability of\linebreak
our approach, as in \cite{lommen2022AutomaticComplexityAnalysis} now we consider so-called \emph{simple cycles}, where
itera\-tions through the cycle can only be done in a unique way.
So the cycle must not\linebreak
have subcycles and there must not be any indeterminisms concerning the next transition to be taken.
Formally, $\mathcal{C} = \{t_1,\ldots,t_n\}\subseteq\TSet$ is a simple cycle if there are pairwise different locations $\location_1,\ldots,\location_n$ such that $t_i = (\location_i, \_, \_, \location_{i+1})$ for $1 \leq i \leq n-1$ and $t_n = (\location_n, \_, \_, \location_1)$.
To handle simple cycles, we \emph{chain} transitions.\footnote{The chaining of a loop $L$ in \Cref{def:chaining} corresponds to $p-1$ chaining steps of a transition $t_L$ via \Cref{Chaining Transitions}, i.e., to $t_L \chain \ldots \chain t_L$.}

\begin{definition}[Chaining (see, e.g., \cite{lommen2022AutomaticComplexityAnalysis})]
	\label{Chaining Transitions}
	Let $t_1,\ldots,t_n \in \TSet$ where $t_i = (\location_i,\linebreak \guard_i, \update_i, \location_{i+1})$ for all $1 \leq i \leq n-1$.
	Then the transition $t_1 \chain \ldots \chain t_n = (\location_1, \guard, \update, \location_{n+1})$ results from \emph{chaining} $t_1,\ldots,t_n$ where
	\[
		\begin{array}{rcl}
			\guard     & = & \guard_1 \, \land \, \update_1(\guard_2) \, \land \, \update_1(\update_2(\guard_3)) \, \land \, \ldots \, \land \, \update_1(\ldots\update_{n - 1}(\guard_n)\ldots) \\
			\update(v) & = & \update_1(\ldots\update_n(v)\ldots) \text{ for all $v \in \VSet$, i.e., $\update = \update_1 \circ \ldots \circ \update_n$.}
		\end{array}
	\]
\end{definition}

Now we want to compute a \emph{local} size bound for the transition $t_n$ w.r.t.\ a simple cycle $\mathcal{C} = \{t_1, \ldots, t_n\}$ where a loop $L$ corresponds to $t_1\chain \ldots \chain t_n$ via $\pi$.
Then a size bound $\Size_L$ for the loop $L$ yields the size bound $\pi \circ \Size_L \circ \pi^{-1}$ for $t_n$ regarding runs through $\mathcal{C}$ starting in $t_1$.
However, to obtain a local size bound $\Size_{t_n}$ w.r.t.\ $\mathcal{C}$, we have to consider runs starting after any entry transition $(\_,\_,\_,\location_i)\in\entry_{\mathcal{C}}$.
Hence, we use $| \, \update_i(\ldots\update_n(\pi(\Size_L(\pi^{-1}(x))))\ldots) \, |$ for any $(\_,\_,\_,\location_i)\in\entry_{\mathcal{C}}$.
In this way, we also capture evaluations starting in $\location_i$, i.e., without evaluating the complete cycle.

\begin{theorem}[Local Size Bounds for Simple Cycles]
	\label{thm:cycles}
	Let $\mathcal{C} = \braced{t_1,\ldots,t_n} \subseteq\TSet$ be a simple cycle and
	let $\Size_L$ be a size bound for a loop $L$ which corresponds to $t_1\chain \ldots \chain t_n$ via a variable renaming $\pi$.
	Then a \emph{local size bound} $\Size_{t_n}$ for $t_n$ w.r.t.\ $\mathcal{C}$ is $\Size_{t_n}(x) = \sum_{1 \leq i \leq n, (\_,\_,\_,\location_i)\in\entry_{\mathcal{C}}} \; | \, \update_i(\ldots\update_n(\pi(\Size_L(\pi^{-1}(x))))\ldots) \, |$.
\end{theorem}
\makeproof{thm:cycles}{
	\begin{myproof}
		We have to show that for all $x\in\VSet$, \Cref{def:local_size_bounds} holds for $\Size_{t_n}(x)$, i.e., for all $\state\in\State$ we have
		\[
			|\state|(\Size_{t_n}(x)) \geq \sup\{ |\state'(x)|\mid \exists \location' \in \LSet, (\_,\_,\_,\location)\in\entry_{\mathcal{C}}.\; (\location,\state) \; (\rightarrow^*_{\mathcal{C}}\circ \rightarrow_{t_n}) \; (\location',\state')\}.
		\]
		Here, the run always has the following form for an $\pret = (\_,\_,\_,\location_j)\in\entry_{\mathcal{C}}$:
		\[
			(\location_j,\state)\; (\rightarrow_{t_j}\circ\ldots\circ \rightarrow_{t_n}) \; (\location_1,\state')\; (\rightarrow_{t_1}\circ \ldots\circ\rightarrow_{t_n})^*\;(\location_1, \state'')
		\]
		So we have to prove that $|\state|(\Size_{t_n}(x))$ is a bound on $|\state'(x)|$ and on all $|\state''(x)|$.
		We have
		\begin{align*}
			|\state|(\Size_{t_n}(x)) & = |\state|\left(\sum\nolimits_{1 \leq i \leq n, (\_,\_,\_,\location_i)\in\entry_{\mathcal{C}}} \; | \, \update_n(\ldots\update_i(\pi(\Size_L(\pi^{-1}(x))))\ldots) \, | \, \right) \\
			                         & \geq |\state| \left(\, | \, \update_j(\ldots\update_n(\pi(\Size_L(\pi^{-1}(x))))\ldots) \, | \, \right)                                                                            \\
			\tag{as every summand is non-negative}
		\end{align*}

		Note that $|\state|(|\,(\update_j\circ ... \circ\update_n\circ
                \pi)(x_i)\,|) \geq |\state \circ \update_j\circ ... \circ\update_n\circ
                \pi| (x_i)$ for all variables $x_i \in \{x_1,\dots,x_d\}$ of the loop $L$.
		Moreover,  $\Size_L(\pi^{-1}(x)) \in \BoundSet$ for all $x \in
                \VSet$.
		Hence, we have
		\begin{align*}
			|\state| \left(\, | \, \update_j(\ldots\update_n(\pi(\Size_L(\pi^{-1}(x))))\ldots) \, | \, \right) & \geq |\state \circ \update_j\circ\ldots\circ\update_n\circ \pi| (\Size_L(\pi^{-1}(x))) \\
			                                                                                                   & = |\state'\circ \pi| (\Size_L(\pi^{-1}(x))).
		\end{align*}
		Since $\Size_L$ is a size bound for the loop $L$, by \cref{Size Bounds of Loops}
		we have $|\state'\circ \pi| (\Size_L(\pi^{-1}(x))) \geq |\state'\circ \pi|(\pi^{-1}(x)) =|\state'| (\pi(\pi^{-1}(x))) = |\state'|(x)$.
		Hence, this proves $|\state|(\Size_{t_n}(x)) \geq |\state'|(x)$.

		Now we prove that $|\state|(\Size_{t_n}(x))$ is a bound on $|\state''(x)|$ if $(\location_1,\state') \; (\rightarrow_{t_1}\circ \ldots\circ\rightarrow_{t_n})^+\;(\location_1, \state'')$ where ``$+$'' denotes the transitive closure.
		By a similar argument as in \cref{lem:lift_size_bounds_loops}, $\pi \circ \Size_L \circ \pi^{-1}$ is a local size bound for $t_1\chain \ldots \chain t_n$ w.r.t.\ $\{ t_1\chain \ldots \chain t_n \}$, i.e., we have
		\begin{align*}
			|\state'| (\pi(\Size_L(\pi^{-1}(x)))) & \geq \sup\braced{|\widetilde{\state}(x)|\mid (\location_1,\state') \; (\rightarrow_{t_1}\circ \ldots\circ\rightarrow_{t_n})^+ \; (\location_1,\widetilde{\state})} \\
			                                      & \geq |\state''|(x).
		\end{align*}
		So in total we have $|\state|(\Size_{t_n}(x)) \geq |\state'| (\pi(\Size_L(\pi^{-1}(x)))) \geq |\state''|(x)$.
	\end{myproof}
}

\begin{example}
	As an example, in the program of \Cref{fig:ITS_solvable_loop}
	we replace $t_1 = (\location_1, x_3 > 0,\linebreak \update_{1}, \location_1)$ by $t_{1a} = (\location_1, \true, \update_{1a}, \location_1')$ and $t_{1b} = (\location_1', x_3 > 0, \update_{1b}, \location_1)$ with a new location $\location_1'$, where $\update_{1a}(v) = \update_1(v)$ for $v \in \{ x_1, x_2 \}$, $\update_{1b}(v) = \update_1(v)$ for $v \in \{ x_3, x_4 \}$, and $\update_{1a}$ resp.\ $\update_{1b}$ are the identity on the remaining variables. Then $\braced{t_{1a},t_{1b}}$ forms a simple cycle and \Cref{thm:cycles} allows us to compute local size bounds
	$\Size_{t_{1b}}$ and $\Size_{t_{1a}}$ w.r.t.\ $\braced{t_{1a},t_{1b}}$, because
        the chained transitions
        $t_{1a} \chain t_{1b} = t_1$ and
$t_{1b} \chain t_{1a}$ both
        correspond to the loop \eqref{WhileExample}.
	They can then be lifted to global size bounds as in \Cref{size bound t1}
	using size bounds for the entry transitions $\entry_{\braced{t_{1a},t_{1b}}} = \{t_0, t_3\}$.
\end{example}

This shows how we choose $t'$ and $\TSet'$ when lifting local size bounds to global ones with \Cref{thm:lift_size_bounds}:
For a transition $t'$ we search for a simple cycle $\TSet'$ such that chaining the cycle results in a twn- or suitable solvable loop and the size bounds of $\entry_{\TSet'}$ are finite.
For all other transitions, we compute size bounds as in \cite{brockschmidt2016AnalyzingRuntimeSize}.

%% file: complexity.tex
\section{\hspace*{-.2cm}Completeness of Size and Runtime Analysis for Programs}
\label{sect:complexity}

For individual loops, we showed in \Cref{thm:completeness} that polynomial runtime bounds
and finite size bounds are computable for all terminating prs loops.
In this section, we discuss completeness  \pagebreak[2] of the size bound technique from the previous section and of termination and runtime complexity analysis for general integer programs.
We show that for a large class of programs consisting of consecutive prs loops, in case of termination we can always infer finite runtime and size bounds.

To this end, we briefly recapitulate how size bounds are used to compute runtime bounds for general integer programs, and show that our new technique to infer size bounds also results in better runtime bounds.
We call $\glo: \TSet \rightarrow \BoundSet$ a \emph{(global) runtime bound} if for every transition $t\in\TSet$ and state $\sigma_0 \in \Sigma$, $|\initial|(\glo(t))$ over-approximates the number of evaluations of $t$ in any run starting in $(\location_0,\state_0)$.
\begin{definition}[Runtime Bound \cite{brockschmidt2016AnalyzingRuntimeSize,Festschrift}]
	\label{def:gloUpperTimeBound}
	A function $\glo: \TSet \rightarrow \BoundSet$ is a \emph{(global) runtime bound}
	if for all $t \in \TSet$ and all states $\initial \in \State$, we have $|\initial|(\glo(t)) \; \geq \; \timeboundterm$.
\end{definition}

For our example in \Cref{fig:ITS_solvable_loop}, a global runtime bound for $t_0$, $t_2$, and $t_3$ is $\glo(t_0)\linebreak
	= 1$ and $\glo(t_2) = \glo(t_3) = x_5$, as $x_5$ is bounded from below by $t_3$'s guard $x_5 > 1$ and the value of $x_5$ decreases by 1 in $t_3$, and no transition increases $x_5$.

To infer global runtime bounds automatically, similar as for size bounds, we first
consider a smaller subprogram $\TSet'\subseteq\TSet$ and
compute \emph{local runtime bounds} for non-empty subsets $\TSet'_>\subseteq\TSet'$.
A local runtime bound measures how often a transition $t\in\TSet'_>$ can occur in a run through $\TSet'$ that starts after an entry transition $\pret\in\entry_{\TSet'}$.
Thus, local runtime bounds do not consider how many $\TSet'$-runs take place in a global run and they do not consider the sizes of the variables before starting a $\TSet'$-run.
We lift these local bounds to global runtime bounds for the complete program afterwards.

\begin{definition}[Local Runtime Bound \cite{lommen2022AutomaticComplexityAnalysis}]
	\label{def:locUpperTimeBound}
	Let $\emptyset\neq\TSet'_>\subseteq \TSet'\subseteq\TSet$.
	$\loc\in\BoundSet$ is a \emph{local runtime bound} for $\TSet'_>$ w.r.t.\ $\TSet'$ if for all $t \in \TSet'_>$, all $\pret\in\entry_{\TSet'}$ with $r = (\location, \_,\_,\_)$, and all $\state \in \State$, we have $|\state|(\loc) \; \geq \; \sup \braced{ n \in \NN \mid \exists\, \initial, (\location', \state').
			\; (\location_0, \initial) \rightarrow_{\TSet}^* \circ \rightarrow_{\pret} \, (\location, \state) \; (\rightarrow_{\TSet'}^* \circ \rightarrow_t)^n \; (\location', \state') }$.
\end{definition}

\begin{example}
	\label{ex:local runtime bounds}
	In \cref{fig:ITS_solvable_loop}, local runtime bounds for $\TSet'_> = \TSet' = \braced{t_1}$ and for $\TSet'_> =\TSet' = \braced{t_4}$ are $\mathcal{RB}_{\braced{t_1}} = x_3$ and $\mathcal{RB}_{\braced{t_4}} =x_1$.
	Local runtime bounds can often be inferred automatically by approaches based on ranking functions (see, e.g., \cite{brockschmidt2016AnalyzingRuntimeSize}) or by the complete technique for terminating prs loops (see \Cref{thm:completeness}).
\end{example}

If we have a local runtime bound $\loc$ w.r.t.\ $\TSet'$, then setting $\glo(t)$ to $\sum_{\pret \in \entry_{\TSet'}}
	\glo(\pret)\cdot (\loc \left[v/\Size(\pret,v) \mid v\!\in\!\VSet \right])$ for all $t\in\TSet'_>$ yields a global runtime bound \cite{lommen2022AutomaticComplexityAnalysis}.
Here, we over-approximate the number of local $\TSet'$-runs which are star\-ted by an entry transition $\pret \in \entry_{\TSet'}$ by an already computed global runtime bound\linebreak
$\glo(\pret)$.
Moreover, we instantiate each $v \in \VSet$ by a size bound $\Size(\pret,v)$ to consider the size of $v$ before a local $\TSet'$-run is started.
So as mentioned in \Cref{sect:integer_programs}, we need runtime bounds to infer size bounds (see \Cref{thm:size_bounds_closed_form} and the inference of global size\linebreak
bounds in \cite{brockschmidt2016AnalyzingRuntimeSize}), and on the other hand we need size bounds to compute runtime bounds.
Thus, our implementation alternates between size bound and runtime bound computations (see \cite{brockschmidt2016AnalyzingRuntimeSize,lommen2022AutomaticComplexityAnalysis} for a more detailed description of this alternation).

\begin{example}
	\label{ex:global_runtime_bound}
	Based on the local runtime bounds in \Cref{ex:local runtime bounds}, we can compute the remaining global runtime bounds for our example.
	We obtain $\glo(t_1) = \glo(t_0)\cdot\linebreak
		(x_3\left[v/\Size(t_0,v) \mid v\in\VSet \right]) + \glo(t_3)\cdot
        (x_3\left[v/\Size(t_3,v) \mid v\in\VSet \right]) = x_3 + x_5^2$ and $\glo(t_4) =
        \glo(t_2)\cdot (x_1\left[v/\Size(t_2,v) \mid v\in\VSet \right]) = x_5\cdot (4\cdot
        x_1 + 2\cdot x_2 + 14\cdot x_5)$.
 \pagebreak[2]	Thus,\linebreak over\-all
	we have a quadratic runtime bound $\sum_{1 \leq i \leq 5}
		\glo(t_i)$.
Note that it is due to our new size bound technique from
\Cref{sect:size_bounds_closed_forms}--\ref{sect:integer_programs} that we obtain
polynomial runtime bounds in this example.
	In contrast, to the best of our knowledge, all other state-of-the-art tools fail to infer polynomial size or runtime bounds for this example.
        Similarly, if one modifies $t_4$ such that instead of $x_1$, $x_4$ is decreased as
        long as $x_4 > 0$ holds, then our approach again yields a polynomial runtime
        bound, whereas none of the other tools can infer finite runtime bounds.
\end{example}

Finally, we state our completeness results for integer programs.
For a set $\mathcal{C} \subseteq \TSet$\linebreak
and $\location, \location' \in \LSet$, let $\location \rightsquigarrow_\mathcal{C} \location'$ hold iff there is a transition $(\location, \_, \_, \location') \in \mathcal{C}$.
We say that $\mathcal{C}$ is a \emph{component} if we have $\location \rightsquigarrow_\mathcal{C}^+ \location'$ for all locations $\location, \location'$ occurring in $\mathcal{C}$, where $\rightsquigarrow_\mathcal{C}^+$ is the transitive closure of $\rightsquigarrow_\mathcal{C}$.
So in particular, we must also have $\location \rightsquigarrow_\mathcal{C}^+ \location$ for all locations $\location$ in the transitions of $\mathcal{C}$.
We call an integer program \emph{simple} if every component is a simple cycle that is
``reachable'' from any initial state.
\begin{definition}[Simple Integer Program]
	\label{def:simple_integer_program}
	An integer program $\IntProgram$ is \emph{simple} if every component $\mathcal{C}
		\subseteq \TSet$ is a simple cycle, and for every entry tran\-sition
                $(\_,\_,\_,\location)\in\entry_{\mathcal{C}}$
and every $\initial\in\State$,
 there is an evaluation $(\location_0,\initial)
 \rightarrow^*_{\TSet} (\location,\initial)$.
\end{definition}

In \cref{fig:ITS_solvable_loop}, $\TSet\setminus\braced{t_0}$ is a component that is no simple cycle.
However, if we remove\linebreak
$t_3$ and replace $t_0$'s guard by $\true$, then the resulting program $\mathcal{P}'$ is simple (but not\linebreak
linear).
A simple program terminates iff each of its isolated simple cycles terminates.
Thus, if we can prove termination for every simple cycle, then the overall program terminates.
Hence, if after chaining,
every simple cycle corresponds to a\linebreak
 linear, unit prs loop, then we can decide termination and infer polynomial runtime and size bounds for the overall integer program.
For terminating, non-unit prs loops, runtime bounds are still polynomial but size bounds can be\linebreak
exponential.
Hence, then the global runtime bounds can be exponential as well.
Note that in the example program $\mathcal{P}'$ above, the eigenvalues of the update matrices of $t_1$ and $t_4$ have absolute value $1$, i.e., $t_1$ and $t_4$ correspond to unit prs loops.
Hence, by \Cref{thm:completeness_integer_programs} we obtain polynomial runtime and size
bounds \pagebreak[2] for $\mathcal{P}'$.

\begin{theorem}[Completeness Results for Integer Programs]
	\label{thm:completeness_integer_programs}

	\vspace*{-.1cm}

	\begin{enumerate}[(a)]
		\item Termination is decidable for all simple \emph{linear} integer programs where after chaining, all simple cycles correspond to prs loops.
		\item Finite runtime and size bounds are computable for all simple integer programs where after chaining, all simple cycles correspond to \emph{terminating}
		      prs loops.
		\item If in addition to (b), all simple cycles correspond to \emph{unit} prs loops, then the runtime and size bounds are \emph{polynomial}.
	\end{enumerate}
\end{theorem}
\makeproof{thm:completeness_integer_programs}{
	\newcommand{\cupdot}{\mathbin{\mathaccent\cdot\cup}}
	\begin{myproof}
		\begin{enumerate}[(a)]
			\item A simple integer program terminates iff every loop corresponding to a simple cycle terminates, where we only consider loops starting from a target location of an entry transition of the simple cycle.
			      In the following, we prove this claim.
			     We define $\mathfrak{S}$ to be the set consisting of all loops that correspond to $t_1 \chain \ldots \chain t_n$ for a simple cycle $\mathcal{C} = \braced{t_1,\ldots,t_n}$ in $\PP$ where there exists an entry transition from $\entry_{\mathcal{C}}$ ending in the start location of $t_1$.

			      If $\initial\in\State$ is a non-terminating initial state for $\PP$, then as $\LSet$ is finite, there exists an $\location\in\LSet$ such that for all $m\in\NN$ there is an $m' \geq m$ and an evaluation $(\location_0,\initial) \rightarrow^* (\location,\state) \rightarrow_{\mathcal{C}}^{m'} (\location,\state')$ for a component $\mathcal{C}$.
			      Thus, the loop $L\in\mathfrak{S}$ which corresponds to $\mathcal{C}$ does not terminate on $\state\in\State$.

			      For the other direction, if $\state\in\State$ is a non-terminating initial state for $L\in\mathfrak{S}$, and $L$ corresponds to $t_1 \chain \ldots \chain t_n$ for a simple cycle $\mathcal{C} = \{t_1,\ldots,t_n\}$ where the start location of $t_1$ is the target location of an entry transition from $\entry_{\mathcal{C}}$, then there exists an evaluation $(\location_0,\state) \rightarrow^* (\location,\state)$ by \Cref{def:simple_integer_program}.
			      Hence, $\PP$ does not terminate for the initial state $\state$.

			      Thus, we can prove termination for every loop in $\mathfrak{S}$ by \Cref{lem:correctness_chaining}(b) and \cite{frohn2020TerminationPolynomialLoops} in order to decide termination of $\PP$.

			\item We can partition the transitions of a simple integer program into three sets -- initial transitions $\TSet_0$, transitions $\TSet_{\mathrm{cyc}}$ which lie on a simple cycle, and transitions $\TSet_1 = \TSet\setminus(\TSet_0\uplus\TSet_{\mathrm{cyc}})$ which connect simple cycles.
			      \paragraph*{Size and Runtime Bounds for $\TSet_0\uplus\TSet_1$:}
			      Note that $\glo(t) = 1$ is a runtime bound for all $t\in\TSet_0\uplus\TSet_1$.
			      A size bound can be obtained by $\Size(t,x) = |\update(x)|$ for all $t = (\location_0,\_,\update,\_)\in\TSet_0$.
			      Moreover, we can obtain a size bound for all $t = (\_,\_,\update,\_)\in\TSet_1$ by $\Size(t,x) = \sum_{\pret \in \entry_{\braced{t}}}|\update(x)| \left[v/\Size(\pret,v) \mid v\!\in\!\VSet \right]$.
			      Note that the set $\entry_{\braced{t}}$ might contain transitions from $\TSet_{\mathrm{cyc}}$.
			      Due to this, we handle transitions in topological order, starting with $\TSet_0$.
			      In this way, we have already inferred finite size and runtime bounds for every entry transition $\pret\in\entry_{\{t\}}$.

			      \paragraph*{Size and Runtime Bounds for $\TSet_{\mathrm{cyc}}$:}

			      Let $\mathcal{C}
				      \subseteq\TSet_{\mathrm{cyc}}$ be a simple cycle.
			      Assume that we already have inferred finite size and runtime bounds for the entry transitions $\entry_{\mathcal{C}}$ of the component $\mathcal{C}$.
			      Then $\glo(t) = \sum_{\pret \in \entry_{\mathcal{C}}}\glo(\pret)\cdot (\mathcal{RB}_{\{t \}} \left[v/\Size(\pret,v) \mid v\!\in\!\VSet \right])$ yields a finite (global) runtime bound for $t\in\mathcal{C}$.
			      Note that we can compute a polynomial runtime bound $\mathcal{RB}_{\{t \}}$ w.r.t.\ $\mathcal{C}$ if $\PP$ and hence the loop corresponding to $\mathcal{C}$ terminates (see \cite{lommen2022AutomaticComplexityAnalysis,hark2020PolynomialLoopsTermination} and \Cref{lem:correctness_chaining}(d)).
			      Similarly, we can infer finite local size bounds via \Cref{thm:cycles}
			      and then lift them to global size bounds $\Size(t,v)$ via \Cref{thm:lift_size_bounds}.
			\item As we only consider unit prs loops, all local size bounds are polynomial (see \Cref{thm:completeness}).
			      By \Cref{thm:completeness}
			      (resp.\ \cite{hark2020PolynomialLoopsTermination}) also all local runtime bounds are polynomial.
			      Note that polynomial bounds are closed under addition, multiplication, and substitution.
			      Thus, all bounds which are constructed in the previous step (b) are polynomial.
		\end{enumerate}
	\end{myproof}
}

In the definition of simple integer programs (\Cref{def:simple_integer_program}),
we required that for every component $\mathcal{C}$ and
every entry tran\-sition $(\_,\_,\_,\location)\in\entry_{\mathcal{C}}$, there is an evaluation $(\location_0,\initial)
\rightarrow^*_{\TSet} (\location,\initial)$ for every
$\initial\in\State$. If one strengthens this by requiring
that one can reach $\location$ from $\location_0$ using only transitions whose guard is
$\true$ and whose update is the identity, then the
class of programs in \Cref{thm:completeness_integer_programs} (a) is
decidable
(there are only $n$ ways to chain a simple cycle with $n$ transitions and checking whether a
loop is a prs loop is decidable by \Cref{Bound on the Period}).

%% file: conclusion.tex
\section{Conclusion and Evaluation}
\label{sect:conclusion}

\paragraph*{Conclusion}
In this paper, we developed techniques to infer size bounds automatically and to use them in order to obtain bounds on the runtime complexity of programs.
This yields a complete procedure to prove termination and to infer runtime and size
bounds for a large class of integer programs.
Moreover, we showed how to integrate
the complete technique into an (incomplete) modular technique for general integer programs.
To sum up, we presented the following new contributions in this  paper:
\begin{enumerate}[(a)]
	\item We showed how to use closed forms in order to infer size bounds for loops with possibly non-linear arithmetic in \cref{thm:size_bounds_closed_form}.
	\item We proved completeness of our novel approach for terminating prs loops (see \cref{thm:completeness}) in \Cref{sect:solvable_loops}.
	\item We embedded our approach for loops into the setting of general integer programs in \cref{sect:integer_programs} and showed completeness of our approach for simple integer programs with only prs loops in \cref{sect:complexity}.
	 \item Finally, we implemented a prototype of our procedure in our re-implementation of the tool \KoAT, written in \tool{OCaml}.
	       It integrates the computation of size bounds via closed forms for twn-loops
               and homogeneous (and thus linear) solvable loops into the complexity
               analysis for general integer programs.\footnote{For a homogeneous solvable loop, the closed form of the twn-loop over $\AA$ that results from its transformation is particularly easy to compute.}
	      \end{enumerate}

To infer local runtime bounds as in \Cref{def:locUpperTimeBound},
\KoAT{}
first applies multiphase-linear ranking functions (see \cite{Festschrift,ben-amram2017MultiphaseLinearRankingFunctions}), which can be done very efficiently.
For twn-\linebreak
loops where no finite bound was found, it then uses the computability of runtime bounds for terminating
twn-loops 
(see \cite{frohn2020TerminationPolynomialLoops,hark2020PolynomialLoopsTermination,lommen2022AutomaticComplexityAnalysis}).
When computing size bounds, \KoAT{}
first applies the technique of \cite{brockschmidt2016AnalyzingRuntimeSize} for reasons of efficiency and in case of exponential or infinite size bounds, it tries to compute size bounds via closed forms\linebreak
as in the current paper.
Here, \tool{SymPy} \cite{sympy} is used to compute Jordan normal forms for the transformation to twn-loops.
Moreover, \KoAT{} applies a local control-flow refinement technique \cite{Festschrift} (using the tool \tool{iRankFinder} \cite{domenech2018IRankFinder}) and preprocesses the program in the beginning, e.g., by extending the guards of transitions with invariants inferred by \tool{Apron}
\cite{jeannet2009ApronLibraryNumerical}.
For all SMT problems, \KoAT{} uses \tool{Z3} \cite{moura2008}.
In
the\linebreak future, we plan to extend the runtime bound inference of \KoAT{} to prs loops and to
extend our size bound computations also to suitable non-linear non-twn-loops.

\paragraph*{Evaluation}
To evaluate our new technique, we tested \tool{KoAT} on the 504 benchmarks for \emph{Complexity of} \tool{C} \emph{Integer Programs}
(\tool{CINT})
from the \emph{Termination Problems Data Base} \cite{tpdb} which is used in the annual \emph{Termination and Complexity Competition (TermComp)} \cite{giesl2019TerminationComplexityCompetition}.
Here, all variables are interpreted as integers over $\ZZ$ (i.e., without overflows).
To distinguish the original version of \KoAT{} \cite{brockschmidt2016AnalyzingRuntimeSize}
from our re-implementation, we refer to them as \tool{KoAT1} resp.\ \tool{KoAT2}.
We used the following configurations of \tool{KoAT2}, which apply different techniques to infer size bounds.

\medskip
\hspace*{-.75cm}
\begin{minipage}{12.5cm}
	\begin{itemize}
		\item[$\bullet\!$] \tool{KoAT2orig} uses the original technique from \cite{brockschmidt2016AnalyzingRuntimeSize} to infer size bounds.
		\item[$\bullet\!$] \tool{KoAT2\,\!+\,\!SIZE} additionally uses our novel approach with \cref{thm:size_bounds_closed_form}, \ref{thm:lift_size_bounds}, and \ref{thm:cycles}.
	\end{itemize}
\end{minipage}

\medskip

The \tool{CINT}
collection contains almost only
examples with linear arithmetic and the existing tools can already solve most
of its benchmarks which are not known to be non-terminating.\footnote{\tool{iRankFinder} \cite{domenech2018IRankFinder} proves non-termination for 119  programs  in \tool{CINT}. \tool{KoAT2orig}
already\linebreak infers finite runtimes for 343 of the remaining $504-119 = 386$ examples in
\tool{CINT}.}
While most complexity analyzers are essentially restricted to programs with linear arithmetic, our new approach also succeeds on programs with \emph{non-linear} arithmetic.
Some programs with non-linear arithmetic could already be handled by
\KoAT{} due to our
integration of the complete technique for the inference of local runtime bounds in
\cite{lommen2022AutomaticComplexityAnalysis}.
But the approach from the current paper increases \KoAT's power substantially for programs
(possibly with non-linear arithmetic) where the values of variables computed in
``earlier'' loops influence the runtime of ``later'' loops (e.g., the modification of our
example from \Cref{fig:ITS_solvable_loop} where $t_4$ decreases $x_4$ instead of $x_1$,
see the end of \Cref{ex:global_runtime_bound}).

Therefore, we extended \tool{CINT} by 15 new typical benchmarks including the
programs in \eqref{WhileExample},
\cref{fig:ITS_solvable_loop},
and the modification of \cref{fig:ITS_solvable_loop} discussed above,
as well as several benchmarks from the literature (e.g., \cite{ben-amram2019TightWorstCaseBounds,ben-amram2008LinearPolynomialExponential}), resulting in the collection \tool{CINT${}^+$}.
For \tool{KoAT2} and \tool{KoAT1}, we used \tool{Clang} \cite{clang} and \tool{llvm2kittel} \cite{falke2011TerminationAnalysisPrograms} to transform \tool{C} into integer programs as in \Cref{sect:integer_programs}.
We compare \tool{KoAT2} with \tool{KoAT1} \cite{brockschmidt2016AnalyzingRuntimeSize} and the tools \tool{CoFloCo} \cite{cofloco2},
\tool{MaxCore} \cite{albert2019ResourceAnalysisDriven} with \tool{CoFloCo} in the backend, and \tool{Loopus} \cite{sinn2017ComplexityResourceBound}.
These tools also rely on variants of size bounds: \tool{CoFloCo} uses a set of constraints to measure the size of variables w.r.t. their initial and final values, \tool{MaxCore}'s size bound computations build upon \cite{DBLP:conf/popl/CousotH78}, and \tool{Loopus} considers suitable bounding invariants to infer size bounds.

\begin{table}[t]
	\makebox[\textwidth][c]{
	  \begin{tabular}{l|c|cc|cc|cc|cc|cc|c|c}
			                                 & $\landau(1)$ &
                  \multicolumn{2}{c|}{$\landau(n)$} & \multicolumn{2}{c|}{$\landau(n^2)$}
                  & \multicolumn{2}{c|}{$\landau(n^{>2})$} &
                  \multicolumn{2}{c|}{$\landau(\mathit{EXP})$} &
                  \multicolumn{2}{c|}{$< \omega$} & 
                  $\mathrm{AVG^+(s)}$ & $\mathrm{AVG(s)}$                               \\
			\hline \tool{KoAT2\,\!+\,\!SIZE} & 26           & 233
                        & (2)                                & 71
                        & (1)                                          & 25
                        & (9)                 & 3                 & (2) & 358 & (14) 
                        & 9.97 
												& 22.88 
												\\
			\hline \tool{KoAT2orig}          & 26           & 232
                        & (1)                                 & 70
                        &                                              & 15
                        &                     & 5                 & (4) & 348 & (5) 
                                                & 8.29 
												& 21.52 
												\\
			\hline \tool{MaxCore}            & 23           & 220
                        & (4)                                 & 67
                        & (1)                                          & 7
                        &                    & 0                 &     & 317 & (5) 
                                                & 1.96 
												& 5.25 
												\\
			\hline \tool{CoFloCo}            & 22           & 197
                        & (1)                                 & 66
                        &                                              & 5
                        &                    & 0                 &     & 290 & (1) 
                                                & 0.59 
												& 2.68 
												\\
			\hline \tool{KoAT1}              & 25           & 170
                        & (1)                                 & 74
                        &                                              & 12
                        &                     & 8                 & (3) & 289 & (4) 
                                                & 0.96 
												& 3.49 
												\\
			\hline \tool{Loopus}             & 17           & 171
                        & (1)                                 & 50
                        & (1)                                          & 6
                        & (1)                 & 0                 &     & 244 & (3) 
                        & 0.40 
												& 0.40 
												\\
	\end{tabular}
	}	\vspace*{.2cm}
	\caption{Evaluation on the Collection \tool{CINT${}^+$}\vspace*{-.4cm}}
	\label{fig:CINT}
     \end{table}

\Cref{fig:CINT} gives the results of our evaluation, where as in \emph{TermComp}, we used a timeout of 5 minutes per example.
The first entry in every cell denotes the number of benchmarks from \tool{CINT${}^+$} for which the tool inferred the respective bound.
The number in brackets only considers the 15 new examples.
The runtime bounds computed by the tools are compared asymptotically as functions which depend on the largest initial absolute value $n$ of all program variables.
So for example, \tool{KoAT2\,\!+\,\!SIZE} proved a linear runtime bound for $231 + 2 = 233$ benchmarks, i.e., $\rc(\state)\in\landau(n)$ holds for all initial states where $|\state(v)|\leq n$ for all $v\in\VSet$.
Overall, this configuration succeeds on 358 examples, i.e., ``$< \omega$'' is the number
of examples where a finite bound on the runtime complexity could be computed by the tool
within the time limit.
``$\mathrm{AVG^+(s)}$'' denotes the average runtime of successful runs in seconds, whereas ``$\mathrm{AVG(s)}$'' is the average runtime of all runs.

Already on the original benchmarks \tool{CINT}, integrating our novel technique for the inference of size bounds leads to the most powerful approach for runtime complexity analysis.
The effect of the new size bound technique becomes even clearer when also considering our new examples which contain non-linear arithmetic and loops whose runtime depends on the results of earlier loops in the program.
Thus, 
the new contributions of the paper are crucial in order to extend automated complexity analysis to larger programs with non-linear arithmetic.

\KoAT's source code, a binary, and a Docker image are available at
\[\mbox{\url{https://koat.verify.rwth-aachen.de/size}.}\]
This website also has details on our experiments, a list and description of the new
examples, and \emph{web interfaces} to run \KoAT's configurations directly online.

%% file: mainArxiv.bbl
\providecommand{\noopsort}[1]{}